\documentclass[12pt]{article}

\usepackage{amsthm}
\newtheorem{definition}{Definition}

\usepackage{mathrsfs}
\usepackage{graphicx,color}
\usepackage{amsfonts}
\usepackage[english]{babel}
\usepackage[latin1]{inputenc}
\usepackage{times}
\usepackage[T1]{fontenc}
\usepackage{amsmath,amssymb}
\usepackage{pgf,pgfarrows,pgfnodes}
\usepackage{amsthm}
\newcommand{\bfzeta}{{\pmb \zeta}}
\DeclareMathOperator{\tr}{Tr}
\usepackage{times}
\usepackage{fullpage}
\usepackage{subfigure}
\title{\vspace{-0.5in} Evaluation and Prediction of Polygon Approximations of Planar Contours for Shape Analysis}

\author{Chalani Prematilake and Leif Ellingson\\ Texas Tech University}

\date{}

\begin{document}

\maketitle

\begin{abstract}
Contours may be viewed as the 2D outline of the image of an object. This type of data arises in medical imaging as well as in computer vision and can be modeled as data on a manifold and can be studied using statistical shape analysis. Practically speaking, each observed contour, while theoretically infinite dimensional, must be discretized for computations. As such, the coordinates for each contour as obtained at k sampling times, resulting in the contour being represented as a k-dimensional complex vector. While choosing large values of k will result in closer approximations to the original contour, this will also result in higher computational costs in the subsequent analysis. The goal of this study is to determine reasonable values for k so as to keep the computational cost low while maintaining accuracy. To do this, we consider two methods for selecting sample points and determine lower bounds for k for obtaining a desired level of approximation error using two different criteria. Because this process is computationally inefficient to perform on a large scale, we then develop models for predicting the lower bounds for k based on simple characteristics of the contours.
\end{abstract}

\section{Introduction}
\label{sec:intr}

The expansion of technology over the past few decades has brought with it tremendous amounts of digital imaging data arising in a variety of fields.  Perhaps most visibly, even standard digital cameras are capable of producing images that are suitable for use in a number of applications related to computer vision, including scene recognition and facial recognition.  Additionally, advances in health-care have led to a rapid growth in medical imaging technology.  Among the many types of medical images are x-rays and computed tomography (CT) scans.  When working with such data, researchers are typically concerned with certain features of an image rather than the entire image, itself.  Often, the feature of interest is the shape of the outline of an object depicted in the image, as discussed in Osborne (2012) and Qiu et al. (2014). 

Originally, researchers analyzed these outlines as finite-dimensional configurations by selecting points on the curves and treating these configurations as the data object.  However, because of variation within the images, these configurations are not provided in the same coordinate systems, as they may differ by rotational, translational, and/or scaling factors.  As such, it is necessary to instead consider the {\em similarity shape} of these configurations.  Bookstein (1978) and Kendall (1984) prominently developed methodology for analyzing such data.  Kendall's methodology views the space of shapes as a manifold, so a statistical analysis requires utilizing tools from differential geometry.  In order analyze shape data, a number of parametric methodologies were developed, as discussed in detail in Kent (1992) and Dryden and Mardia (1998).  

Inspired in part by shape analysis, nonparametric approaches to statistics on manifolds were developed by Bhattacharya and Patrangenaru (2003, 2005) and, independently, Hendriks and Landsman (1998) based on ideas from Fr\'echet (1948) and Ziezold (1977).  Subsequently, there have been many papers that have contributed to the literature in this area.  Among them are Huckemann and Ziezold (2006), Bhattacharya (2008), Dryden et al. (2008), Balan et al. (2009), Bandulasiri et al. (2009),  Brombin and Salmaso (2009), Amaral and Wood (2010), Huckemann et al. (2010), Huckemann (2012), Jung et al. (2012), and Osborne et al. (2013). While a number of the papers describe methodologies for general manifolds, many of these focus on specific manifolds arising in a given application.  Many of these methodologies are described in a discussion paper by Bhattacharya and Patrangenaru (2013), and recent monographs by Bhattacharya and Bhattacharya (2012) and Patrangenaru and Ellingson (2015).

Within the past 20 years or so, though, researchers have shifted their focus towards analyzing the outlines in images as continuous oubjects. For instance, Grenander (1993) considered shapes as points on an infinite dimensional space. Furthermore, a manifold model for direct similarity shapes of planar closed curves, which was first suggested by Azencott (1994), was pursued in Azencott et al. (1996), and further detailed by Younes (1998, 1999).  This area gained additional ground the turn of the millennium, as more researchers began studying shapes of planar closed curves, as in Sebastian et al. (2003).  Klassen et al. (2004), Michor and Mumford (2004), and Younes et al. (2008) follow the methods of Small (1996) and Kendall by defining a Riemannian structure on a shape manifold. Klassen {et al.} (2004) compute an intrinsic sample mean shape. Subsequently, Mio et al. (2007), Srivastava et al. (2005), Mio et al. (2004), Joshi  et al. (2007), and, most recently, Kurtek et al. (2012) have explored the similarity shape of closed curves modulo reparametrizations for a chosen Riemannian metric.

However, despite these frameworks being developed for analyzing shapes of continuous curves, discretization of the curves is unavoidable for computations.  Researchers typically approach this problem by sampling from the curves at an arbitrary number of points.  This can be problematic because, if the number of points chosen is too small, then the curves may not be sufficiently well approximated.  On the other hand, though, if too many points are used, then the computational cost associated with the subsequent analysis will be needlessly increased.  This is especially burdensome when a given application requires efficient analysis, as is often the case when working with imaging data.  As such, it is important to discretize in a way that balances the approximation error and computational cost.  While introducing nonparametric methodology for analyzing shapes of contours, Ellingson et al. (2013) briefly discussed this approximation problem.  However, the focus was primarily on inferential procedures.

In this paper, we will build upon the work of Ellingson et al. (2013) by developing a framework for finding a lower bound for the number of points needed when discretizing for adequately approximating the original outlines.  Additionally, we will develop regression models for predicting these lower bounds based on various geometric features of the outlines.  The paper will be organized as follows.

We first discuss about Kendal's shape space which is also known as the direct similarity shape space, and then specifically about similarity shape analysis for planar contours. Then we discuss about the discretization of contours and importance of polygon approximation over $k$-ad. Then we describe the two parameterizations that we use to approximate polygons. Next we talk about the two decision criteria of approximating lower bounds for $k$ at a given error threshold. We use both decision criteria under each of two parameterizations and find four different ways of approximating lower bounds for each observation. Since this method has to use for individual observations, it is time consuming therefore we introduce linear regression models to predict those lower bounds using characteristics of the original observations as predictors. Finally we discuss the predictability and validity of the regression models we derived.
 
\section{Similarity Shape Analysis for Finite Planar Configurations}
\label{sec:sim}

To begin, we must consider in the notion of similarity shape in somewhat more detail.  As such, we will describe the approach set forth by Kendall (1984) in which a configuration is traditionally represented as a $k$-ad, which is a set of $k$ ordered, labelled points of interest called landmarks.  Two $k$-ads of points in the plane,
${\bf z_1}=(z_1^1, \dots, z_1^k), {\bf z_2}=(z_2^1, \dots, z_2^k) \in \mathbb C^k$, are said to have the same direct similarity shape if they differ only by translation, rotation, and/or scale.  The mathematical definition of direct similarity shape is thus defined as follows. Translations are filtered out by centering the $k$-ad ${\bf z}=(z^1, \dots, z^k)$ to
$ \bfzeta = (\zeta^1, \dots, \zeta^k)$, where $\zeta^j = z^j- \overline{z}, \forall j =1, \dots, k$. 

\begin{figure}[!ht]
	\begin{center}
		\includegraphics[scale=0.5]{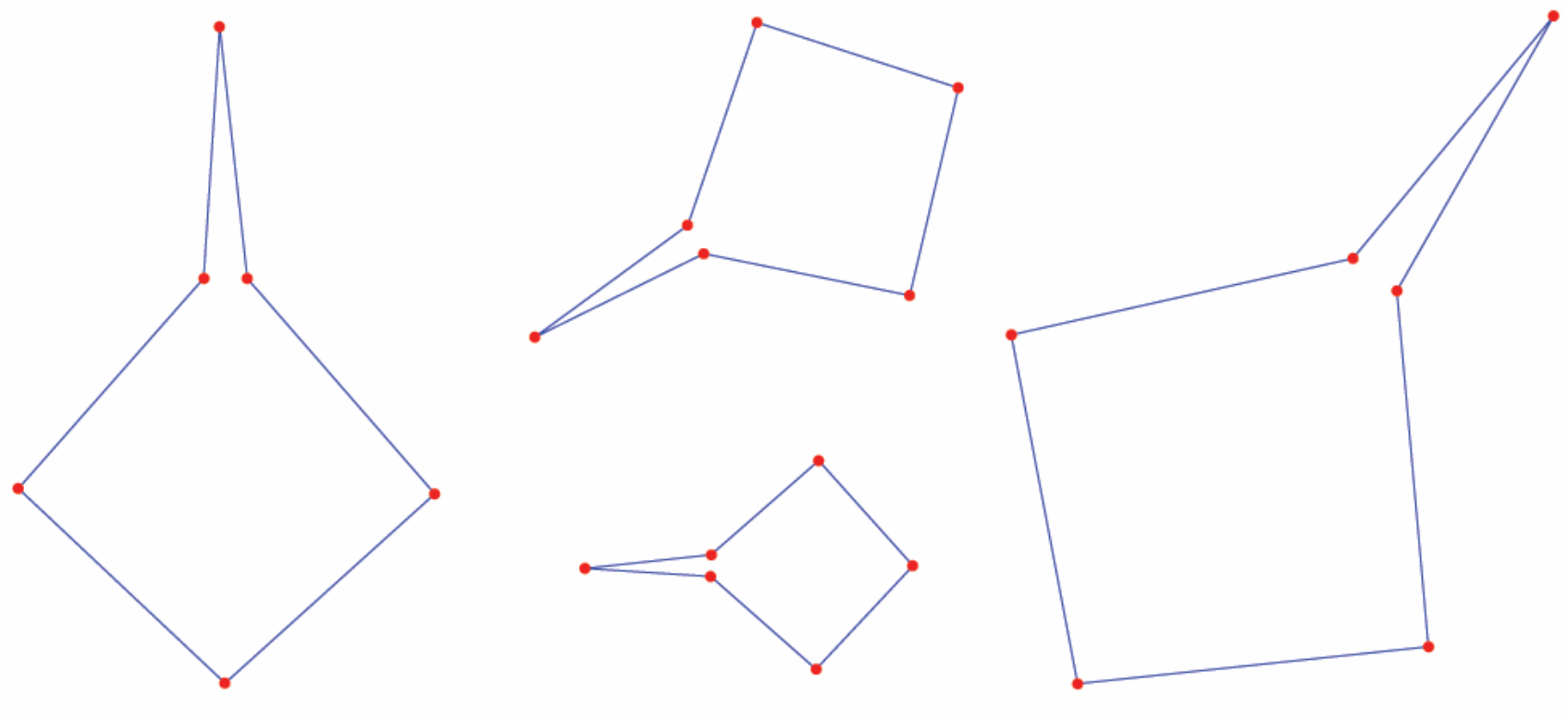}
		\caption{Four configurations that all have the same Kendall shape, differing only in rotation, translation, and scale from each other.}
		\label{f:sameshape}
	\end{center}
\end{figure}

Unfortunately, rotation and scale are not so easily filtered out.  Instead, the direct similarity shape, or Kendall shape of $\bfzeta$ is defined using an equivalence class.  Indeed, the Kendall shape $[{\bfzeta}]$ of $\bfzeta$ is the orbit under rotation and scaling of $\bfzeta.$ That is,

\begin{equation} \label{planarshape} [\bfzeta] =  \{\lambda e^{i\theta} {\bfzeta}: -\pi< \theta \leq
\pi, \lambda > 0 \},
\end{equation}
where $\lambda$ is a scaling factor and $\theta$ is an angle of rotation about the origin.  As a result, the four configurations shown in Figure \ref{f:sameshape}, despite differing from each other, all have the same Kendall shape.  

By virtue of this definition, shape data can be more difficult to work with quantitatively than standard multivariate data.  Indeed, the space of all such shapes, denoted by $\Sigma_2^k$, can be identified with the projective space $\mathbb CP^{k-2}$, which is a manifold, meaning that traditional statistical techniques for Euclidean data cannot be directly used. One approach for working with data on this space is via the Veronese-Whitney embedding, as discussed in Bhattacharya and Patrangenaru (2005), Bandulasiri et al. (2009), and Amaral et al. (2010).

\section{Similarity Shape Analysis of Planar Contours}
\label{sec:pcont}

As stated in the introduction, we will utilize the formulation of similarity shape for planar contours as presented in Ellingson et al. (2013), which defined general methodology for data analysis on Hilbert manifolds with a focus on the explicit example of shapes of planar contours.  As such, a contour $\tilde \gamma$ is defined to be the range of a piecewise differentiable function $\gamma$ that is parametrized by arclength where $\gamma: [0, L] \rightarrow \mathbb C,$ such that $\gamma(0) = \gamma(L)$ and $\gamma$ is one-to-one on $[0, L)$.  That is to say that the contour $\tilde{\gamma}$ is  non-self-intersecting. For convenience, we will subsequently identify the contour with the function $\gamma$ and consider only those contours said to be {\em regular}, which means that there is a unique point $z_0 = argmax_{z \in { \gamma}}\|z-z_{{\gamma}}\|,$ where $ z_{\gamma}$ denotes the center of mass of $\gamma$.  In other words, there is a unique point on the contour that is furthest from the center of mass.  

Just as in the the finite-dimensional case, the direct similarity shape $[\gamma]$ of a contour ${\gamma}$ consists of all contours differing from $\gamma$ by only rotation, translation, and/or scale.  We can again filter out translations by centering the contour as follows: $\gamma_0 = \gamma -\bar{z}_{ \gamma} = \{z - \bar{z}_{ \gamma}, z \in  \gamma\}.$ $\gamma_0$ is called a centered contour. Similarly to Section 1, we define the direct similarity shape of $\gamma$ to be the orbit under rotation and scaling of $\gamma_0$.  Symbolically, this can be expressed as:
\begin{equation*} [\gamma] =  \{\lambda e^{i\theta} {\gamma_0}: -\pi< \theta \leq \pi, \lambda > 0 \},
\end{equation*}

where $\lambda$ and $\theta$ are as defined in \eqref{planarshape}.  The space of the shapes, $\Sigma_2^{reg}$, is a Hilbert manifold. To analyze data lying on this space, Ellingson et al. (2013) proposed using the Veronese-Whitney (VW) embedding of this space into the space of Hilbert-Schmidt operators, which is a linear space.  The formula for the  embedding $j$ is given as follows: $j([\gamma]) = \frac{1}{\|\gamma\|^2}\gamma \otimes \gamma,$ where $[\gamma]$ denotes the shape of $\gamma$. The distance $\rho$ induced via $j$ is the Frobenius norm on the space of Hilbert-Schmidt operators. Let $A, B \in P({\bf H}) $. Then,
\begin{equation} \label{rho2}
\rho^2(A,B) = \tr{((j(A)-j(B))\otimes(j(A)-j(B)))}
\end{equation}

\section{Contour Discretization}
\label{sec:discret}

Ideally, we could conduct data analysis directly on the infinite-dimensional space using the preceding results, but because contours are typically not analytical functions, it is necessary to discretize them when performing computations.  Since contours are functional data, we could appeal to techniques described in Ramsay and Silverman (2005) to approximate the contours using a finite number of basis functions of some kind.  Alternatively, though, we can appeal to the fact that digital imaging data is naturally discrete.  However, the number of pixels comprising a contour changes from observation to observation.  In Kendall's framework, this would mean that observation $j$ would be represented as a $k_j$-ad, so the shapes would lie in different spaces, making a statistical analysis impossible as is.  Because of this, we need to choose a constant number $k$ of points, called landmarks, across observations.  At the same point, we want our discretization to appropriately represent the original contour.

As presented in Dryden and Mardia (1998), there are three traditional landmark selection methods: anatomical, mathematical and pseudo. Anatomical landmarks, so named due to its prevalence in medical imaging, require an expert to locate biologically important features on the configuration. For example, if we want to reconstruct a human face, we have to select points to represent important features, such as the eyes, mouth, and nose. Mathematical landmarks are chosen by considering geometrical properties of the contour. For example, we may only want to choose points with relatively high or low curvatures to be landmarks. Both of these approaches typically involve choosing a small number of landmarks, resulting in a fairly low dimensional representation of the original configuration and may not be sufficient for approximating full contours. 

Pseudolandmarks are points constructed without considering any specific geometrical or anatomical property. Perhaps the most commonly used example of these are equally-spaced landmarks, which are constructed on the object so that the physical distance between each pair of consecutively chosen points is the same on the original contour. Another type, as considered in Ellingson et al. (2013), is random landmark selection, where the points are chosen randomly along the contour.  A related type of landmark to these is a semilandmark, which are often referred to as sliding landmarks, as defined in Bookstein (1997).

While the use of pseudolandmarks or semilandmarks typically results in a relatively high dimensional representation of the contour, it is unfortunately rare for researchers to devote much thought to the choice of $k$.  Typically, researchers simply choose some arbitrary number, such as multiples of 50, for $k$.  However, in order to preserve all of the characteristics of the contour so as to minimize the error, the approximation of an object has to be done carefully. On the other hand, because higher dimensional approximations require higher computational costs, it is advantageous to not have too high of a value for $k$.  As such, we seek to find a lower bound for $k$ for a given contour so as to find a balance between these competing factors.

\subsection{Polygon Approximation}
\label{sec:polygon}

Unfortunately, though, evaluating approximations can be tough using $k$-ads because there is a fundamental difference in nature between a finite-dimensional $k$-ad and an infinite-dimensional contour.  Due to this, we want to discretize in a way that we can appeal to the finite-dimensional techniques while also remaining able to quantify approximation error. To solve this problem, we propose to use the representation introduced by Ellingson et al. (2013), which suggested to approximate contours as polygons.  To do this, sampling points are obtained by evaluating the contour $\gamma$ at $k$ times, which we will call sampling times. The linear interpolation of the sampling points yields the $k$-gon $z$. It follows, then, that $z$ is a one-to-one piecewise differentiable function that can be parametrized by arclength. 

As such, $z$ can be evaluated at any number of times. Indeed, for $s \in (0,1)$, the $k$-gon under arc length parameterization can be expressed as follows:
\begin{equation} \label{k-gon}
z(sL_k) =
\begin{cases}
(t_2 -s L_k) z(0) + sL_k z(t_2) &  0 < sL_k \leq t_2\\
(t_j -sL_k) z(t_{j-1}) + (sL_k - t_{j-1}) z(t_j) &  t_{j-1} < sL_k \leq t_j \\
(L_k -sL_k) z(t_{k}) + (sL_k - t_{k}) z(0) &  t_{k} < sL_k < L_k
\end{cases}
\end{equation}
for $j=3,\dots,k$. Here, $L_k$ denotes the length of the $k$-gon, $z(t_j)$ is the $j$th ordered vertex, where $t_j \in [0, L_k)$, and $z(t_1)=z(0)=z( L_k)$.   Physically, this parametrization can be viewed as traversing the contour at constant speed, which we may take for convenience to be unitary, so the distance traveled along the path of the contour from $s_j$ to $s_{j+1}$ depends only on $s_{j+1}-s_j$.  

Another useful representation of contours that we will consider is a curvature parametrization, where we instead traverse a constant amount of curvature per change in interval.  As such, the distance traveled along the path of the contour depends on $s_j$ in addition to $s_{j+1}-s_j$.  That is, if we denote the infinitesimal changes in arc-length, absolute curvature, and distance around the contour, respectively, by $dL$, $d\kappa$, and $dz$, then the speed at which the contour is traversed with respect to absolute curvature is
$
\frac{dz}{d\kappa}=\frac{dz}{dL} \cdot \frac{dL}{d\kappa}.
$

Since $\frac{dz}{dL}$ is constant under the arc-length parametrization, if it is assumed to be unitary, then the speed is equal to $\frac{dL}{d\kappa}$.  That is, we travel along the contour more slowly where the magnitude of the curvature is higher than we do where it is near zero.  This speed conversion allows us to utilize \eqref{k-gon} to obtain $k$-gon approximations under the curvature parametrization, as well.

The space of these $k$-gons is dense in the space of contours, which will allow us to use various methods for comparing our approximations to the original observations.  It should be noted that, in general, the center of mass of a $k$-gon differs from that of its respective $k$-ad because the latter consists only of the sampling points.  This impacts the shapes of these configurations since they are defined in terms of the centered configurations.

The sampling points, which serve as vertices of the polygon (though some may be collinear), are analogous to landmarks in the finite-dimensional setting.  As such, the landmark selection methods described above are suitable approaches for selecting sampling points.  For the purposes of this paper, we will consider sampling points chosen to retain either equal spacing or equal curvature. The first sampling point will always be the one furthest from the center of mass of the contour, which is unique since we consider only regular contours.

\begin{figure}[!ht]
	\begin{center}
		\includegraphics[scale=0.25]{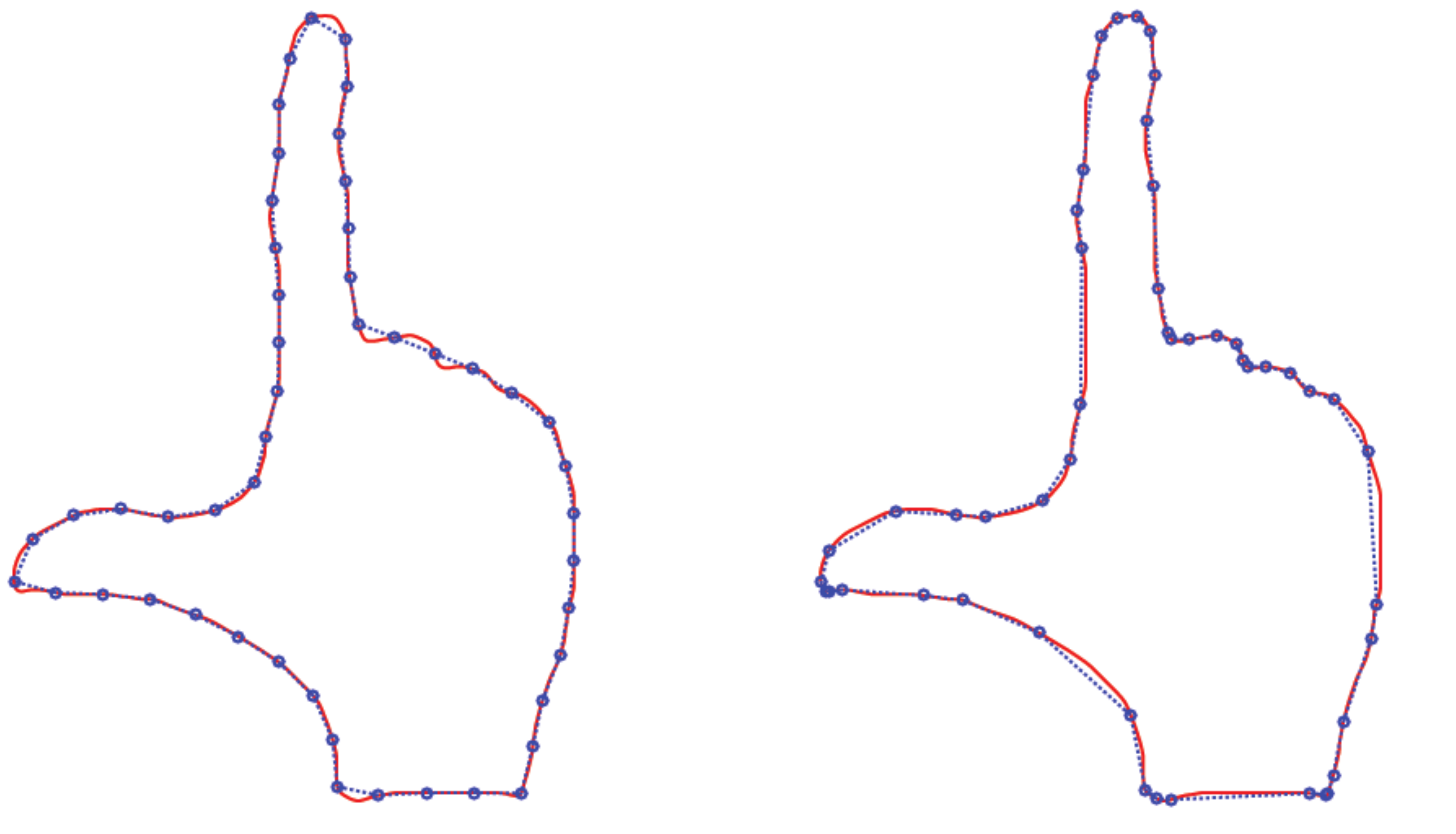}
		\caption{Choosing 50 points along the contour using two different parameterizations.}
		\label{f:2param} 
	\end{center}
\end{figure}

As an illustration of these methods, consider the example depicted in Figure \ref{f:2param}.  Sampling points on the left $k$-gon are equally spaced, whereas the total absolute curvature between each consecutive pair of sampling points is equal in the right $k$-gon. Note that the left $k$-gon misses a number of features, such as around the tip of the finger and the knuckles.  This is in contrast to the figure on the right, which, though sampling points are clustered together where curvature is extreme, the $k$-gon better captures the details the the equally-spaced sampling points do not.

\section{Evaluating Adequacy of Approximations}
\label{sec:dc}

\begin{figure}[!ht] \label{f:kLplots}
	\begin{center}
		\includegraphics[scale=0.7]{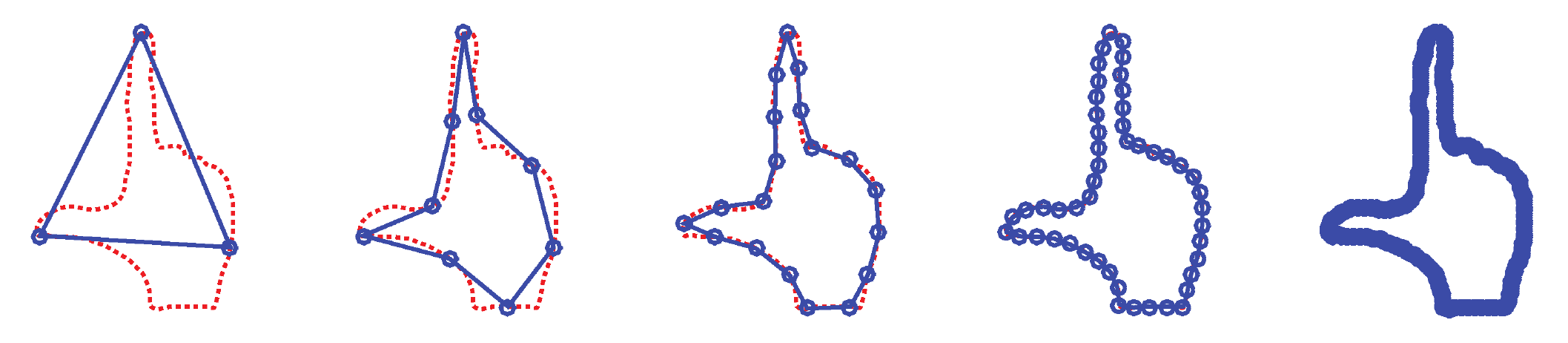}\\
		\caption{Change of shape due to approximation as $k$ increases; approximated polygons for $k=4, 10, 20, 50, \mbox{ and } 250$ in blue on original contours in red}. 
	\end{center}
\end{figure}

Because discretization occurs when obtaining a digital representation of a contour, we identify the discretized digital contour $\gamma_K$ with the ideal contour $\gamma$ and, as such, treat it as the original observation. That is, we assume it has $K$ ordered points that trace the contour when the adjacent points are connected. However, not all $K$ of the points may contain relevant information about the shape of the contour.  At the same time, however, As such, we want to use a polygon, as defined above, with $k$ vertices, for some $4\leq k < K$, to approximate the original observation.  The key to this, then, is determining an appropriate value of $k<<K$ so that the $k$-gon closely resembles the original contour.  To illustrate the importance of the choice for $k$, Figure \ref{f:kLplots} shows polygon approximations with various values of $k$ for the contour of a hand gesture.

As a result of the $k$-gon not being coincident with the original contour, we expect differences between the characteristics of the original contour and the approximated polygon. One such difference is in the lengths of original contour and the approximated polygon. Another characteristic that quantifies the difference is the induced distance between the shapes of the $k$-gon and the original contour in the shape space.    Although we may expect other changes, such as the displacement of the center of mass due to the approximation, in this study we will focus only on the above two changes. We seek to balance the loss of information due to these approximation errors and the increased computational cost for the subsequent analyses associated with higher values of $k$.

To do this, we focused on choosing a value for $k$ by specifying a maximal level of approximation error in terms of both the length and the distance.  Because choosing a number of sampling points greater than $k$ will result in approximation error less than the desired level, the value of $k$ that attains the error threshold is, in fact, a lower bound with respect to that criterion and threshold.  

For our first criterion, we choose $k$ by fixing an upper bound for the relative error of lengths.  Because of the linear interpolations used in obtaining the $k$-gons, for $k'>k$, $L_{k'} \geq L_k$, so the relative error with respect to $K$ will always be positive.  The lowest value of $k$ that satisfies the length criterion will be denoted as $kL$.  It is defined more formally as follows.

\begin{definition} \label{kL-def}
For a given observation with $K$ points, for each $4\leq k < K$ and for a prespecified, small error threshold, $0 < E < 1$, there exists at least one $k$ such that $\frac{L_K-L_{k}}{L_K} \leq E$ where $L_k=\sum_{j=2}^{k+1}\left\|z(t_j)-z(t_{j-1})\right\|$ and $z(0)=z(t_{k+1})$. The smallest $k$ from this set is $kL$.
\end{definition}

For our second criterion, we will utilize the distance between the shapes of the original contour and the approximations in the shape space.  To normalize the distances, we will utilize the fact that, because the shape space is compact, the greatest possible distance under the VW embedding between any two shapes is $\sqrt{2}$.  We will again place an upper threshold on this criterion to choose our lower bound for $k$, which we will denote by $kD$.  

\begin{definition}
For a given observation with $K$ points, for each $4\leq k < K$ and for a prespecified, small error threshold, $0 < E < 1$, there exists at least one $k$ such that $\rho\left(z_k,\gamma_K\right)<\sqrt{2}E$ where $\rho$ is the distance between the approximated $k$-gon, $z_k$, and the original contour, $\gamma_K$. The smallest $k$ from this set is $kD$.
\end{definition}

Before we present an example to illustrate these concepts, we will first explain some notation that will be used throughout the remainder of this paper.  $k_{LA}, k_{DA},k_{LC}$, and $k_{DC}$ denote the lower bounds for number of sampling points. $k$ represents the number of sampling points. The first subscript represents the decision criteria; $L$ for length and $D$ for distance. The second subscript represents the parameterization; $A$ for arclength parameterization and $C$ for curvature based parameterization. Therefore, $k_{LA}$ is the lower bound of $k$ using the length based decision criterion under arclength parameterization.

\begin{figure}[ht!]
\includegraphics[scale=0.27]{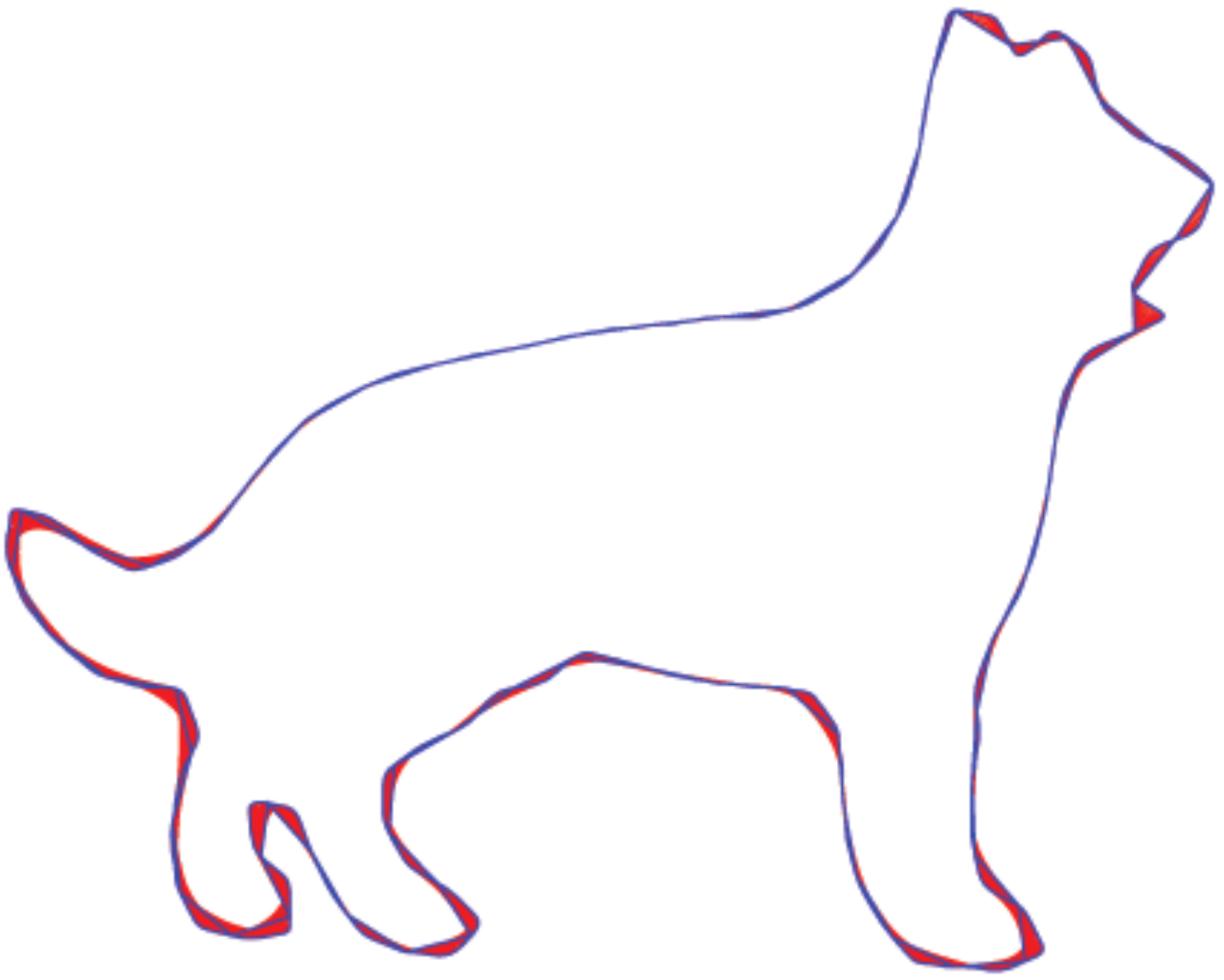}\includegraphics[scale=0.2]{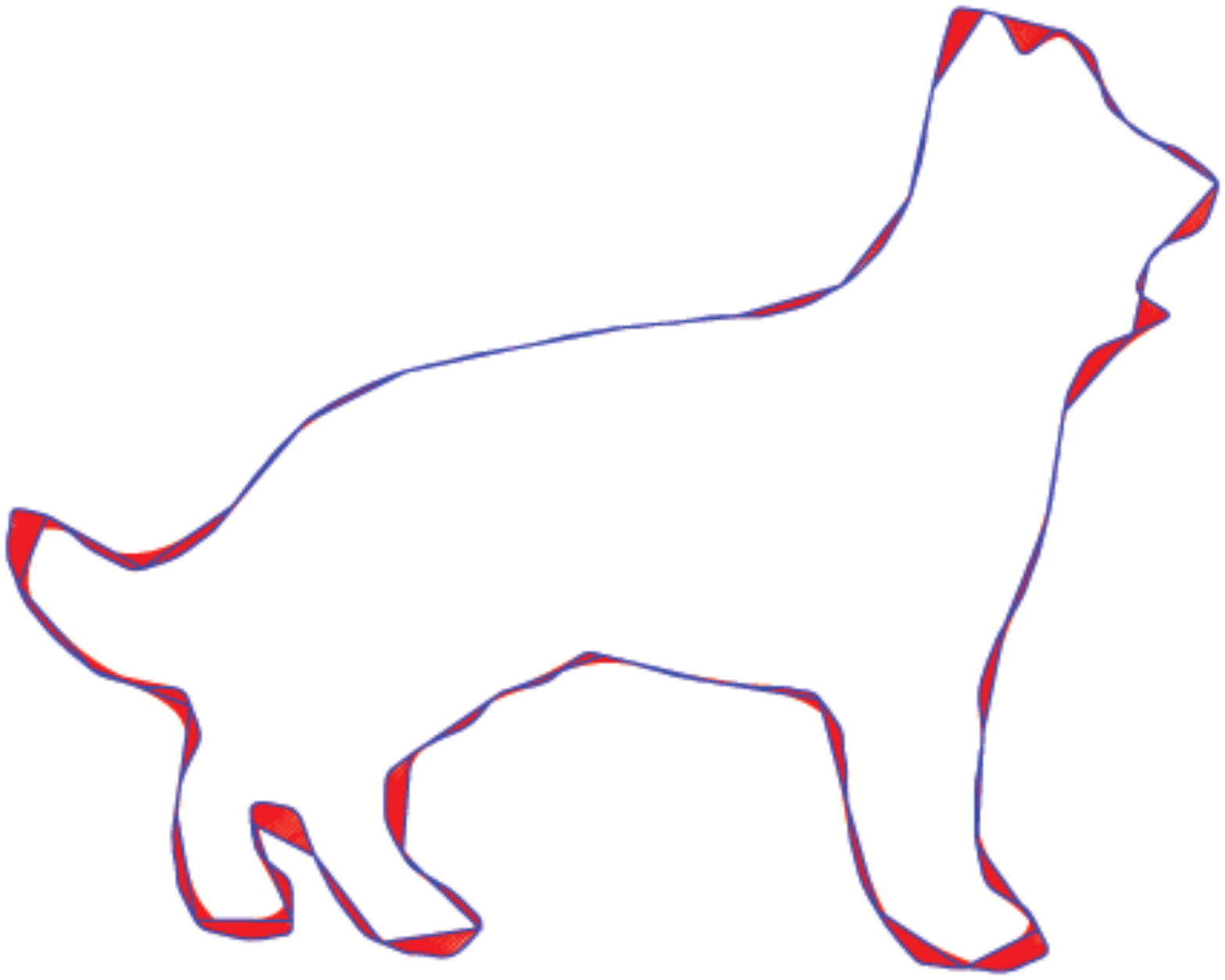}\includegraphics[scale=0.27]{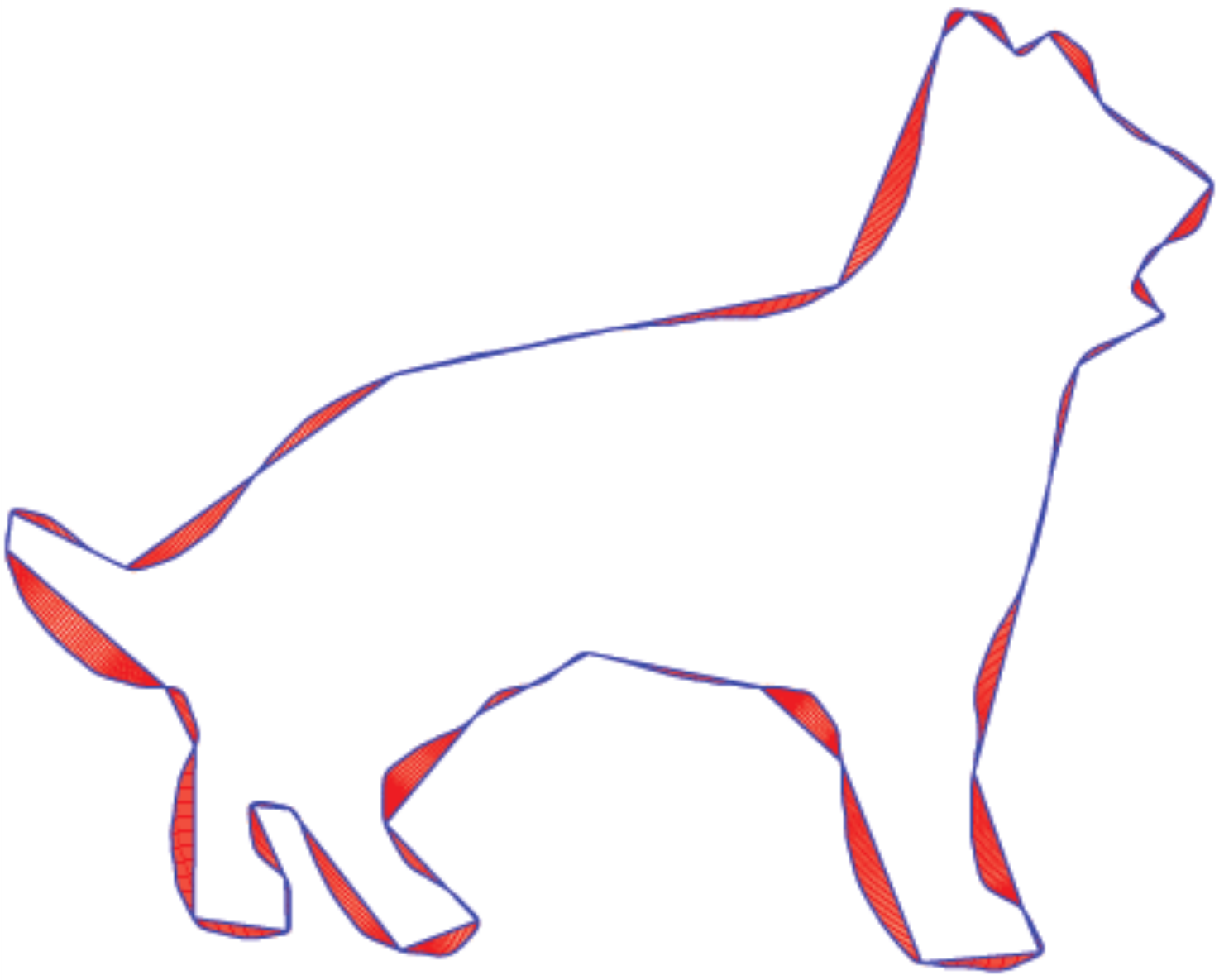}\includegraphics[scale=0.2]{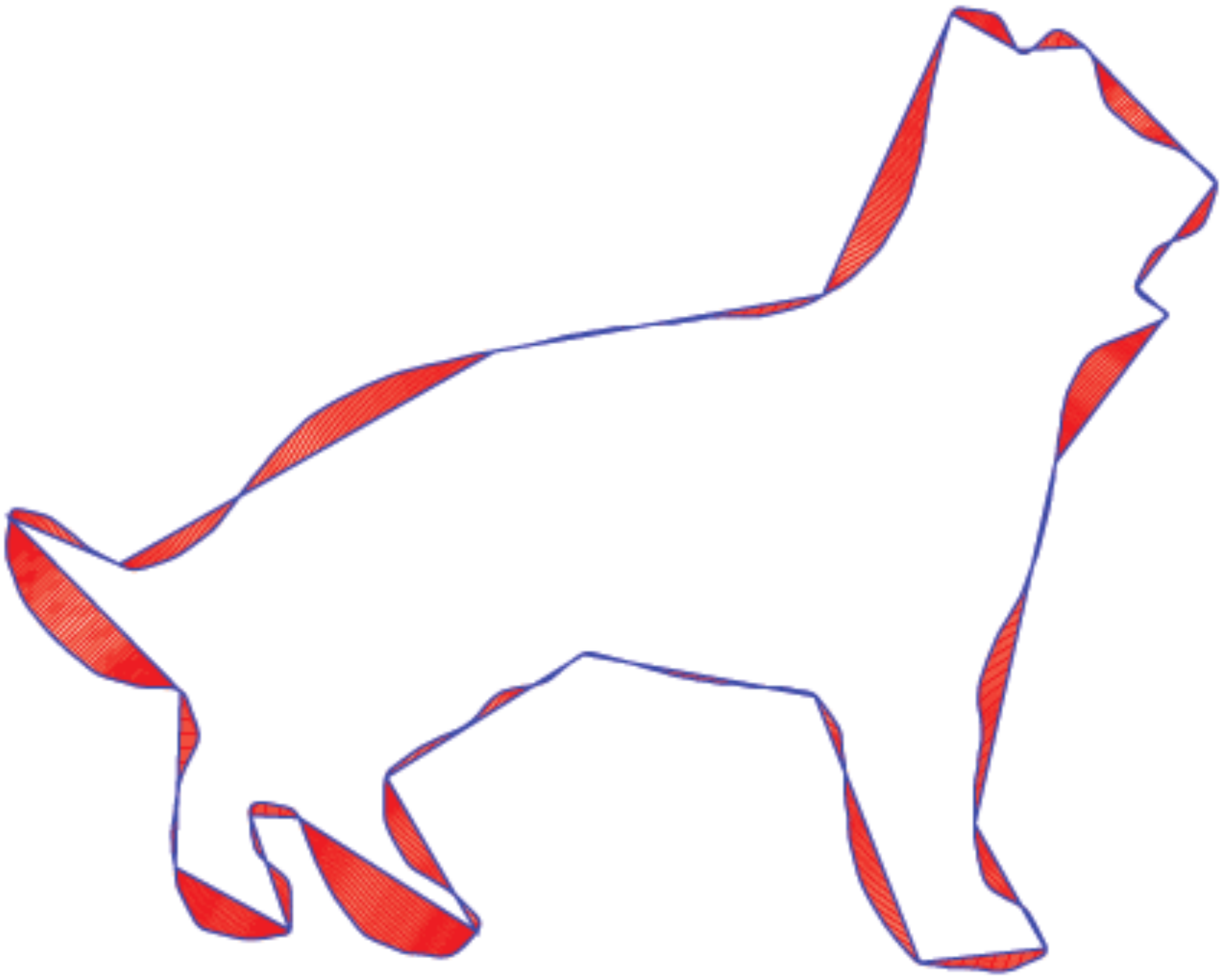}\\
\includegraphics[scale=0.27]{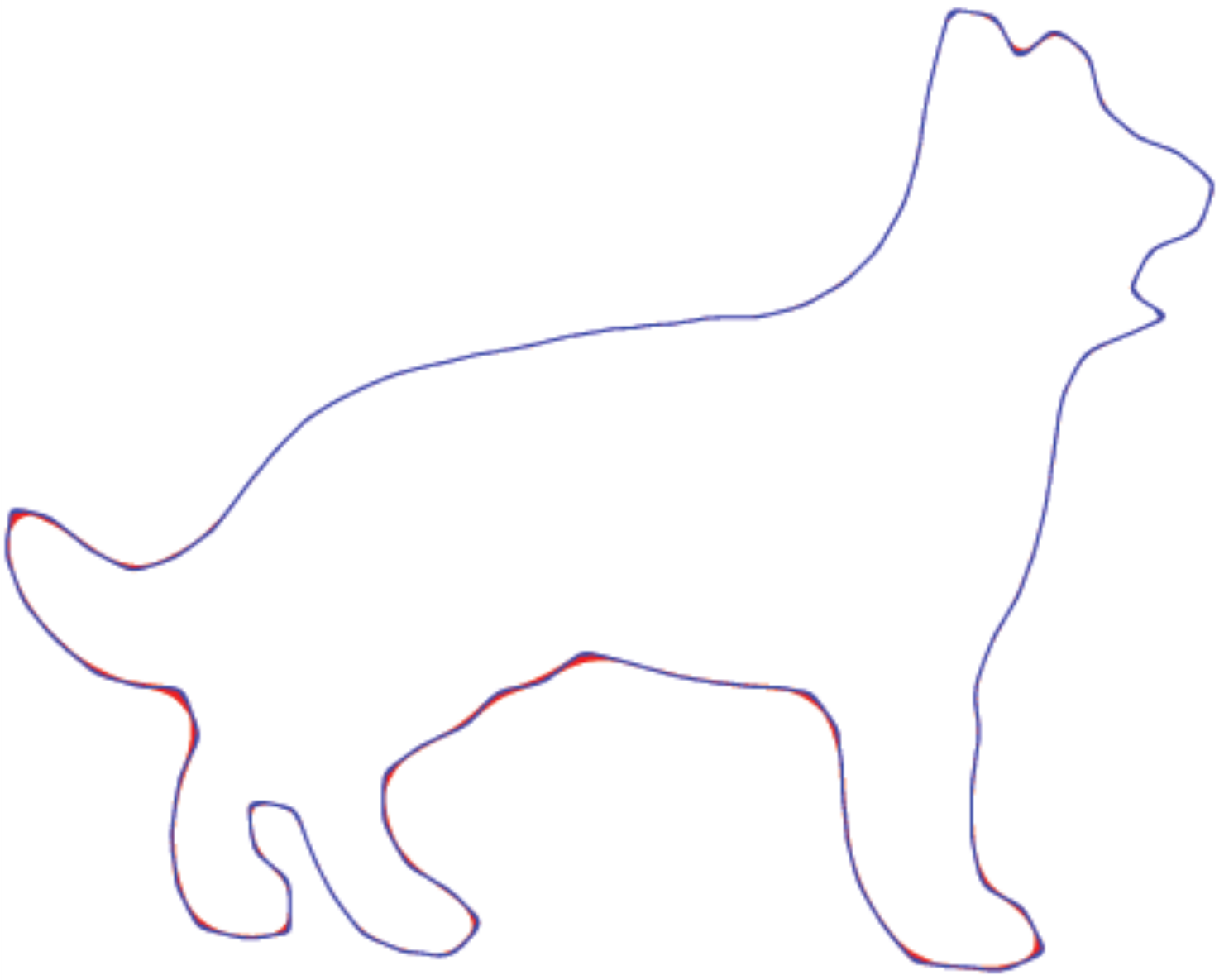}\includegraphics[scale=0.2]{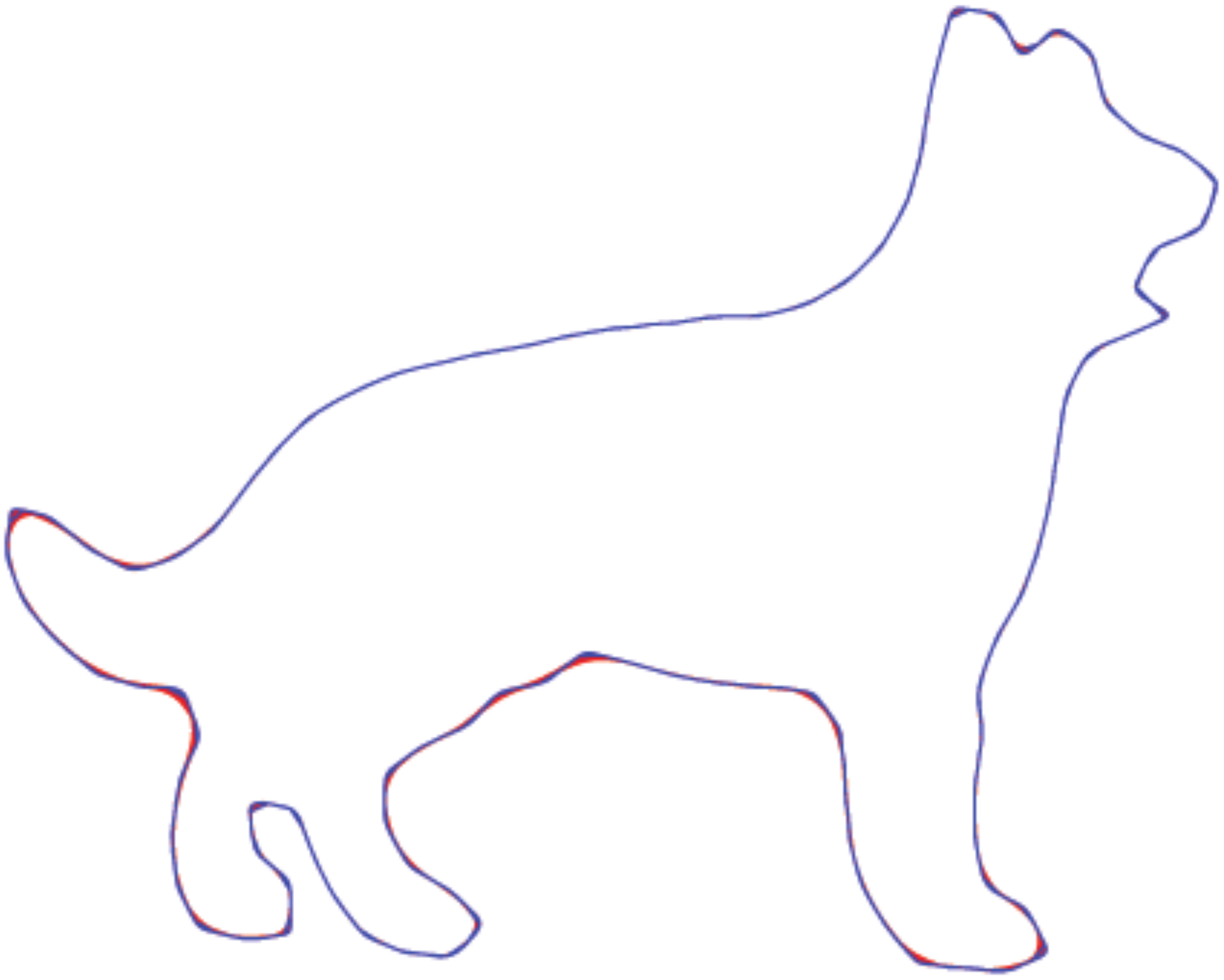}\includegraphics[scale=0.2]{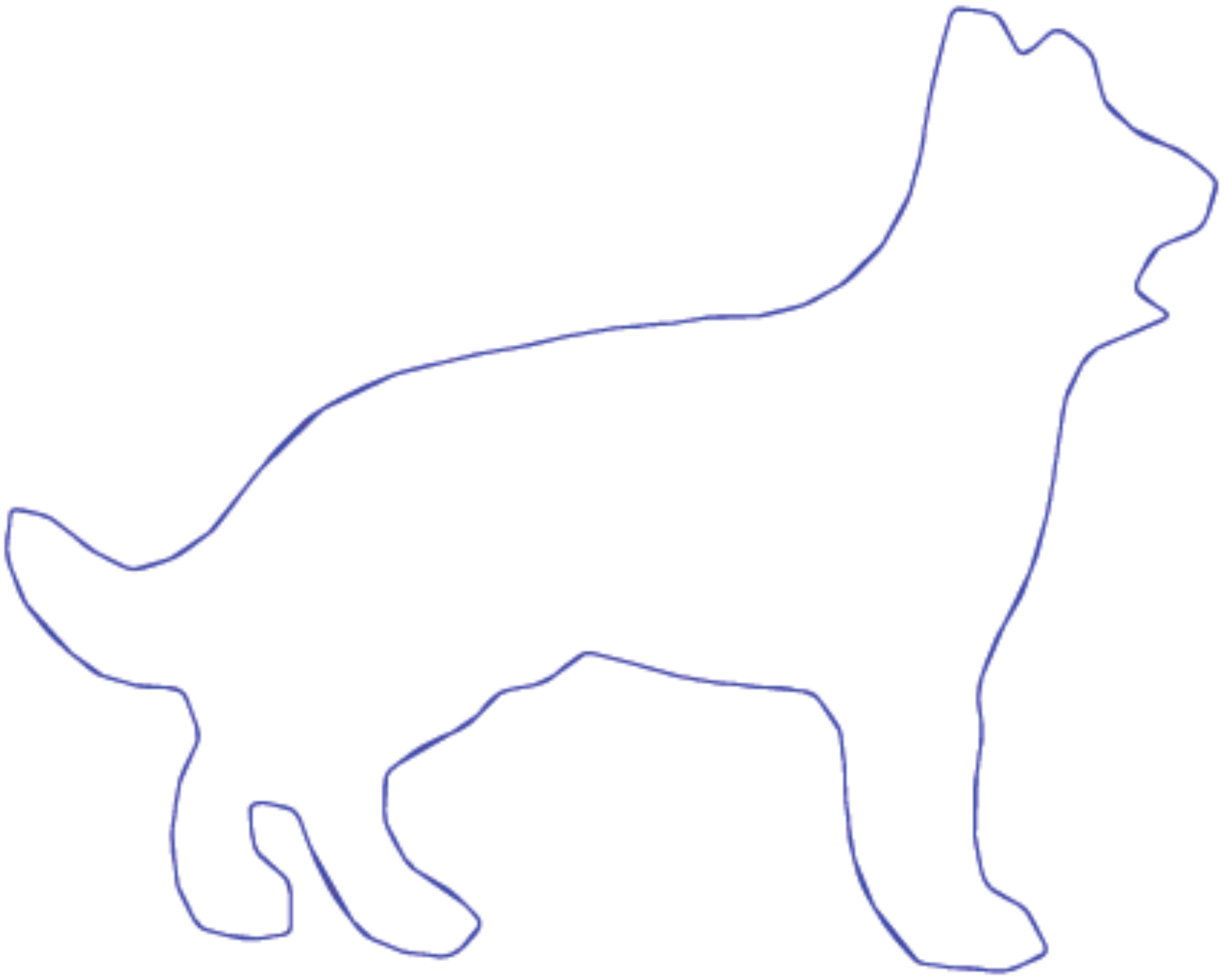}\includegraphics[scale=0.27]{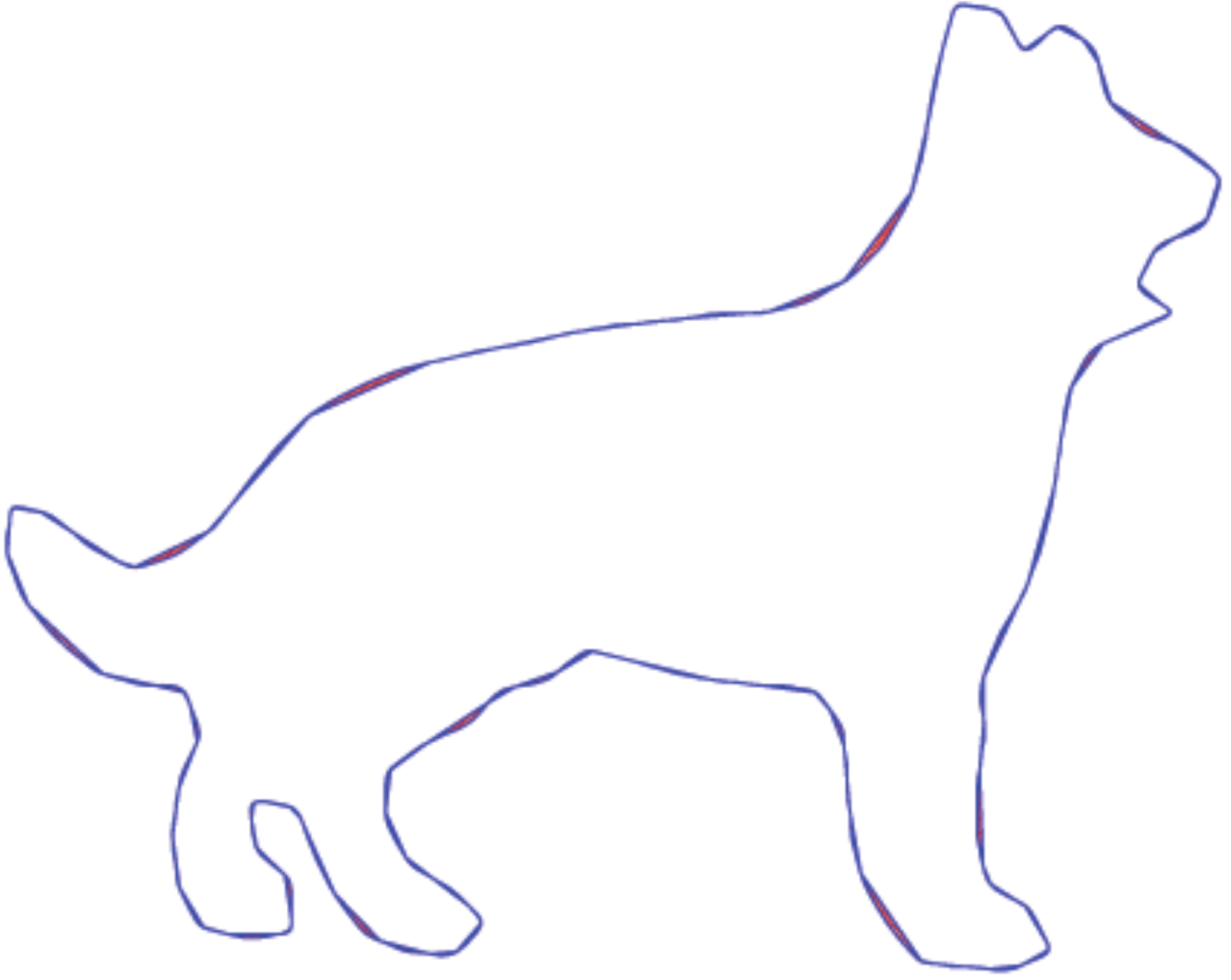}
	\caption{Approximated $k$-gons: $k_{LA}=66$, $k_{DA}=43$, $k_{LC}=36$, $k_{DC}=30$ at $E=0.05$ in the first row and $k_{LA}=303$, $k_{DA}=211$, $k_{LC}=209$, $k_{DC}=133$ at $E=0.005$ in the second row \label{f:dog-example}  }
\end{figure}

Consider an outline of a picture of a dog (Figure \ref{f:dog-example}). The full discretized contour consists of $K=1214$ points. The dog's mouth, ears, tail, and three of its legs are clearly visible. Our goal is to reconstruct the contour with a smaller number of sampling points, $k$. We use the two decision criteria to approximate a lower bound for $K$ under each of the two parameterizations. 

\begin{figure}[ht!]
\center{
	\begin{minipage}[b]{.45\linewidth}
		\includegraphics[scale=0.22]{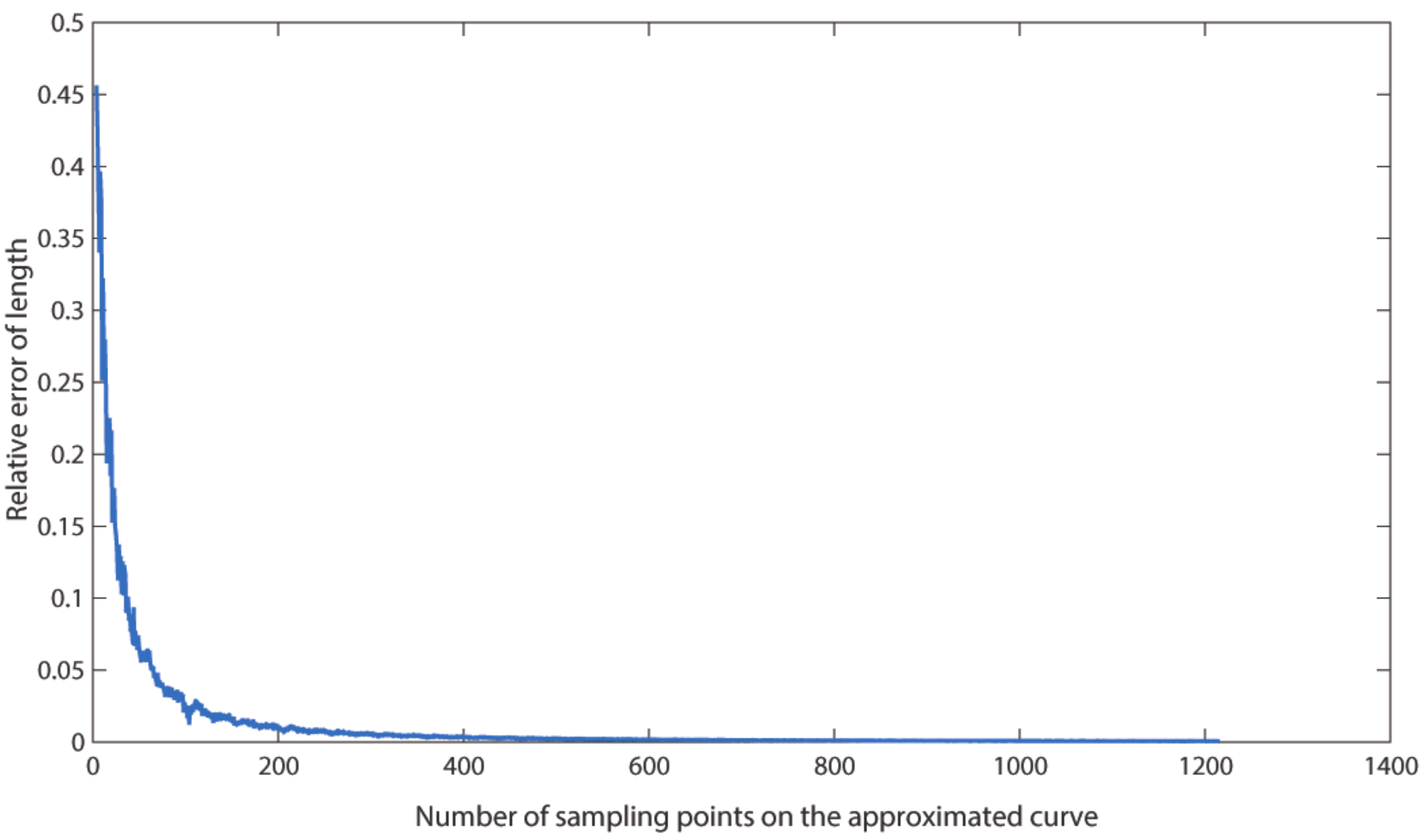}
	\end{minipage}
	\begin{minipage}[b]{.45\linewidth}
		\includegraphics[scale=0.22]{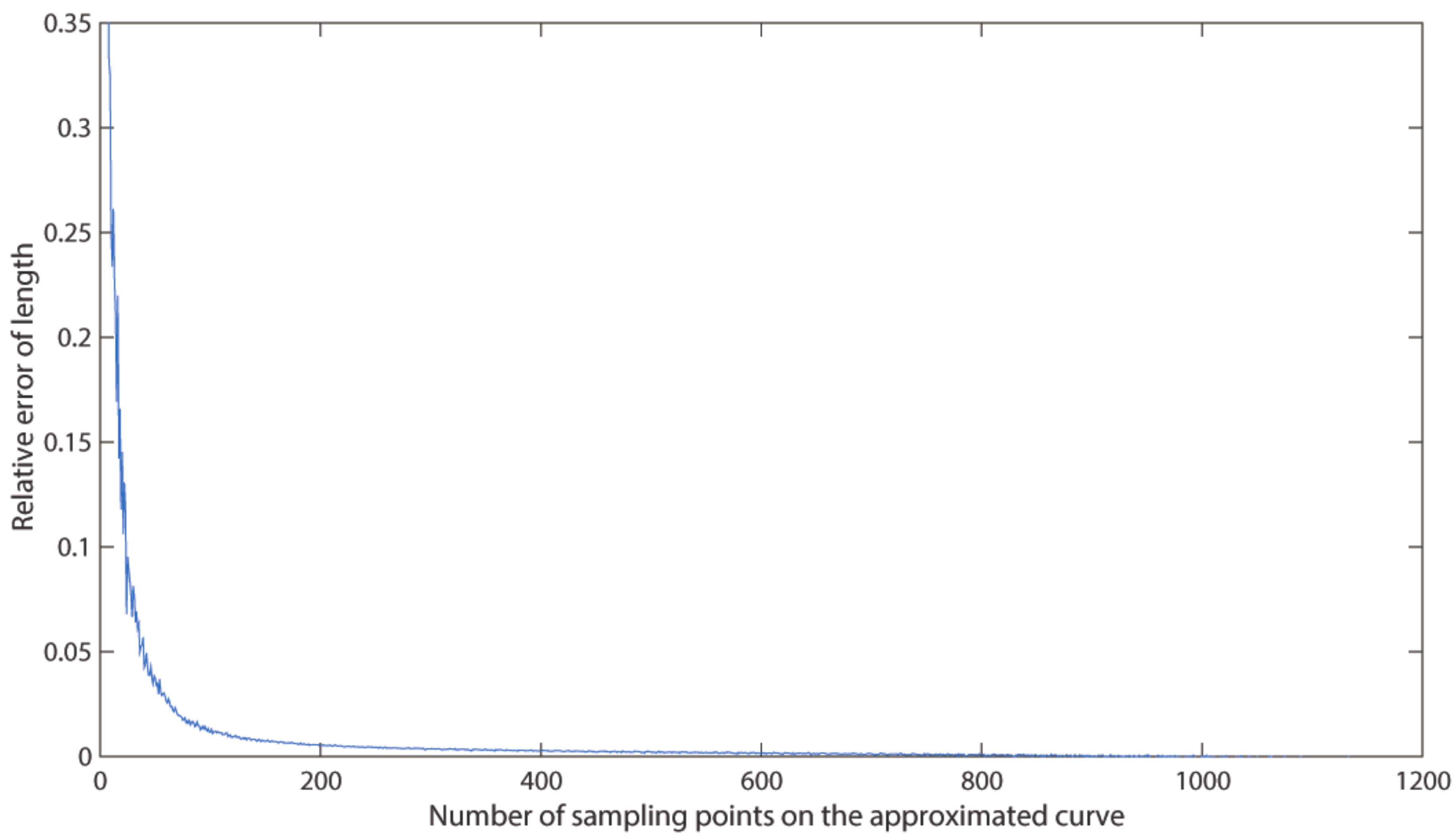}
	\end{minipage}
		\caption{Relative error in length due to approximation vs $k$; left: Arclength parameterization, right: Curvature parameterization \label{f:kL-example}}	
		}
\end{figure}

\begin{figure}[ht!]
\center{
	\begin{minipage}[b]{.45\linewidth}
		\includegraphics[scale=0.22]{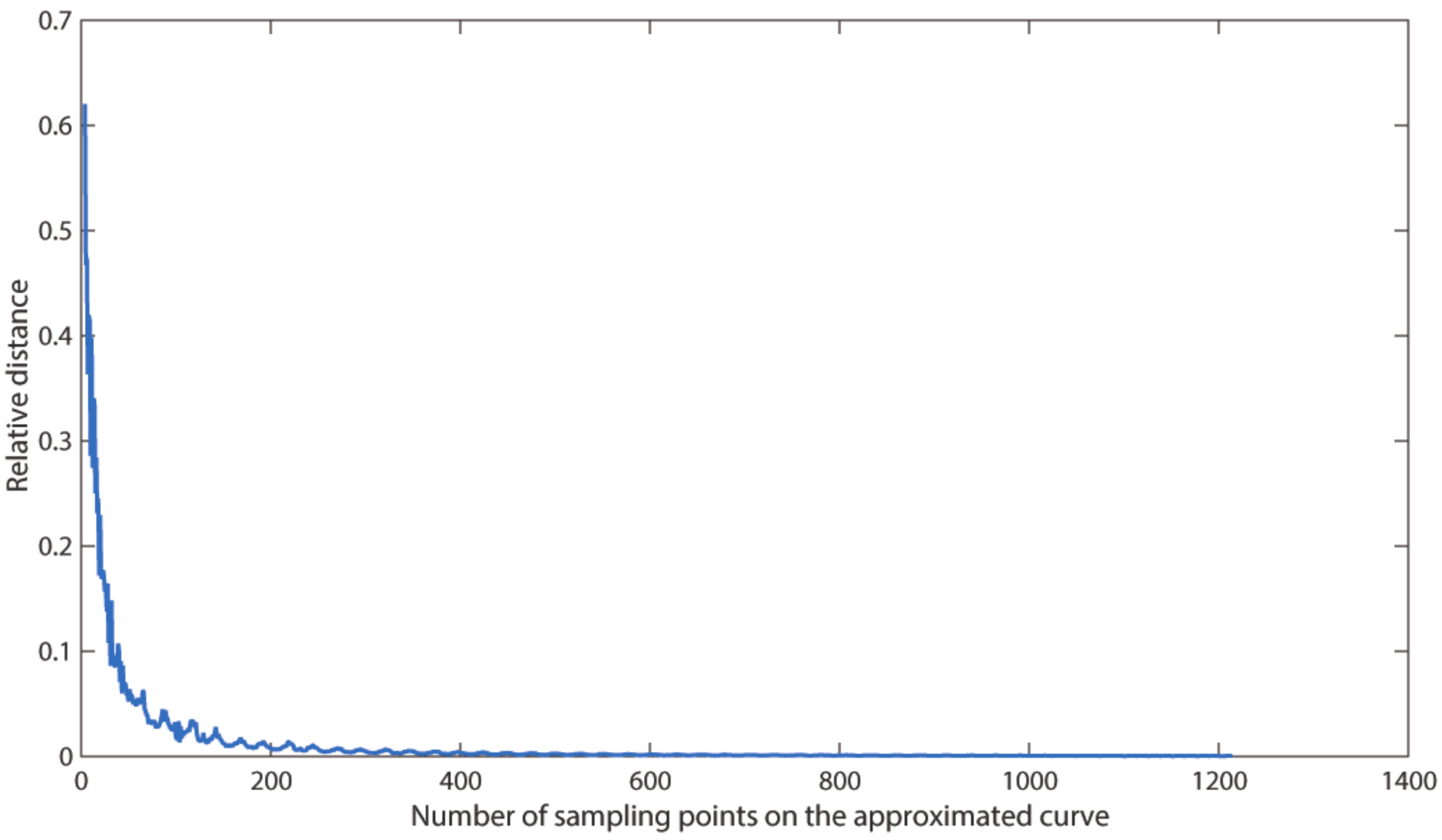}
	\end{minipage}
	\begin{minipage}[b]{.45\linewidth}
		\includegraphics[scale=0.22]{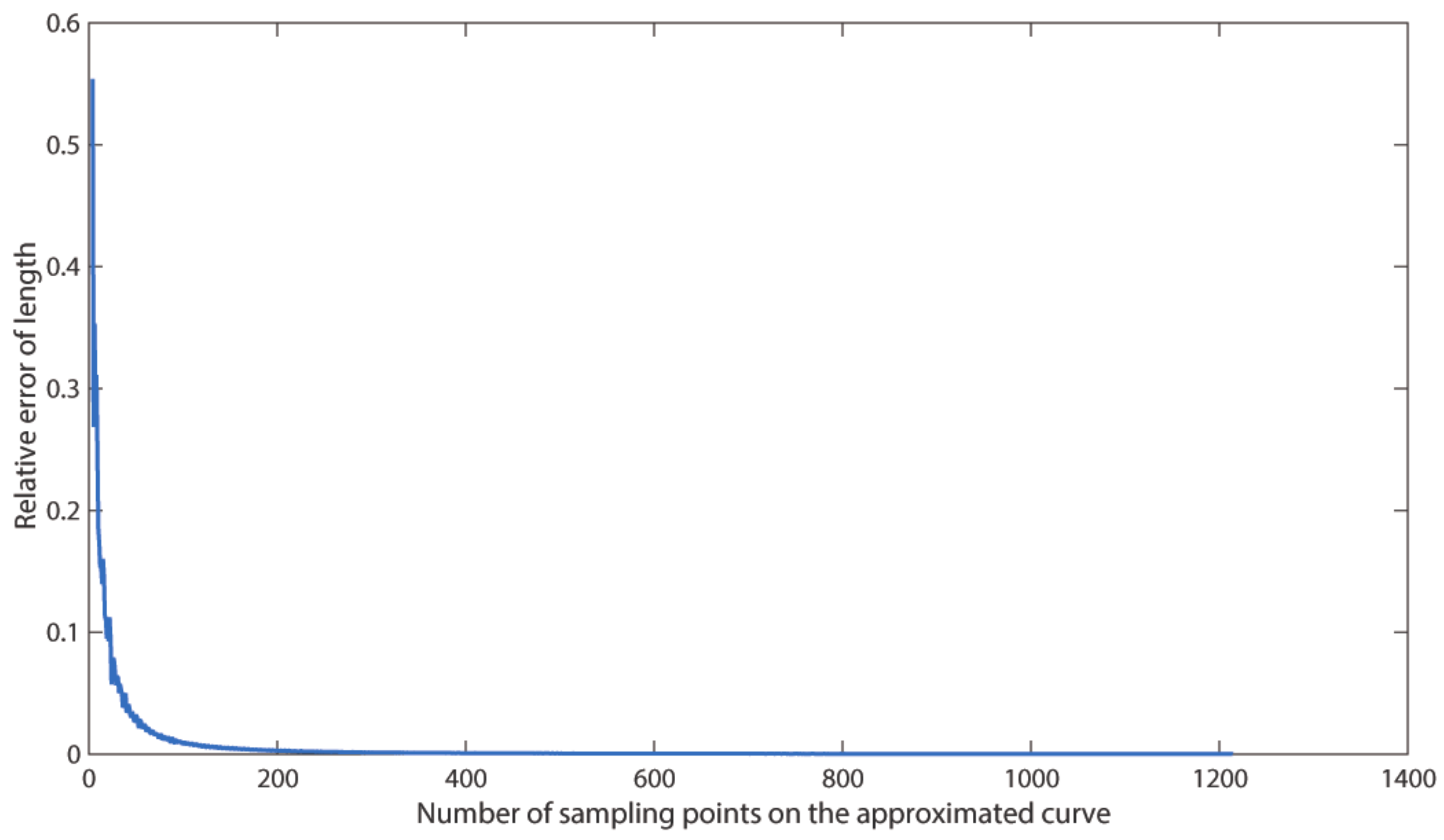}
	\end{minipage}
			\caption{Relative distance between the original contour and the approximated contour vs $k$; left: Arclength parameterization, right: Curvature parameterization  \label{f:kD-example}}
			}
\end{figure}

Figures \ref{f:kL-example} and \ref{f:kD-example} show the relative error due to approximation for all $k \mbox{ from } 4 \mbox{ to } K$ for the length and distance criteria, respectively. The two plots in each figure show the results under the two parameterizations. In all cases, we see that relative error decreases with a decreasing rate as $k$ increases.  As a function of $k$, the relative distance decreases more rapidly than the relative difference in lengths, so, for a given threshold $E$, $k_{DA}$ and $k_{DC}$ are smaller than $k_{LA}$ and $k_{LC}$, respectively.  If we choose $E=0.05$, then for this contour, $k_{LA}=66, k_{DA}=43, k_{LC}=36$, and $k_{DC}=30$. If we choose a smaller $E$, for example $E=0.005$, then for this contour, $k_{LA}=303, k_{DA}=211, k_{LC}=209$, and $k_{DC}=133$. The resulting $k$-gons are shown in Figure \ref{f:dog-example}. Visually inspecting these polygons reveals that all of them contain nearly all the characteristics of the original contour despite being of substantially lower dimension. The $30$-gon at $E=0.05$ under the curvature parameterization for the distance criteria is slightly different from the original contour, as one might expect due to the drastic dimension reduction in that case. At $E=0.005$, curvature parameterization captures almost all the characteristics that captured by arclength parameterization at a comparably smaller $k$.

\section{Smoothing the contours}

\begin{figure}[ht!]
	\center{\includegraphics[scale=0.4]{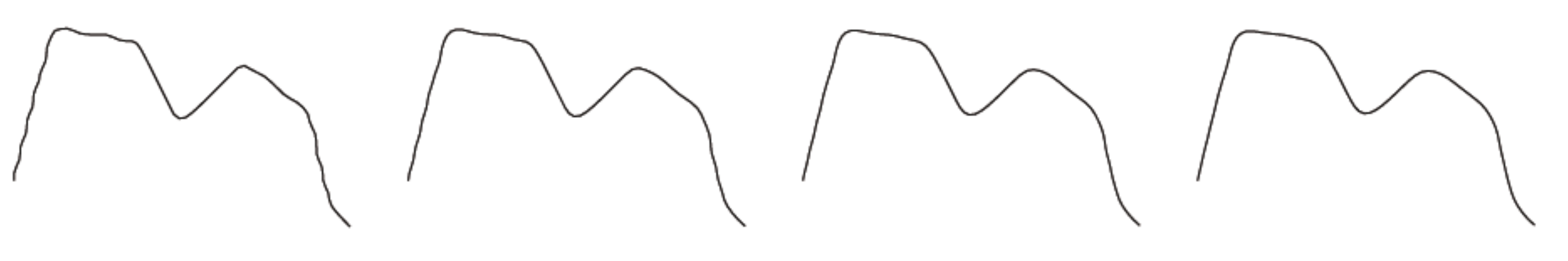}}
	\caption{A portion of the contour of a dog after smoothing the curve once, twice, three times, and four times, respectively. \label{f:dog-smooth}}
\end{figure}

As a result of the discretization inherent to digitization, many contours have small fluctuations that may not be visible when viewing the entire contour, but are noticeable when zooming in on only portions of them.  This is illustrated in the first curve shown in Figure \ref{f:dog-smooth}.  While these fluctuations do not substantially impact the lengths of the contours, they have a tremendous impact on the absolute curvature, both locally and globally.  As such, it is vital to smooth the contours prior to finding approximations.  Here, we chose to use a moving average smoother several times in order to choose an adequate amount of smoothing.  To demonstrate the impact of the moving average smoother, we have plotted a portion of the contour of a dog under successive implementations of the moving average smoother in Figure \ref{f:dog-smooth}.

\begin{figure}[ht!]
\center{
	\includegraphics[scale=0.25]{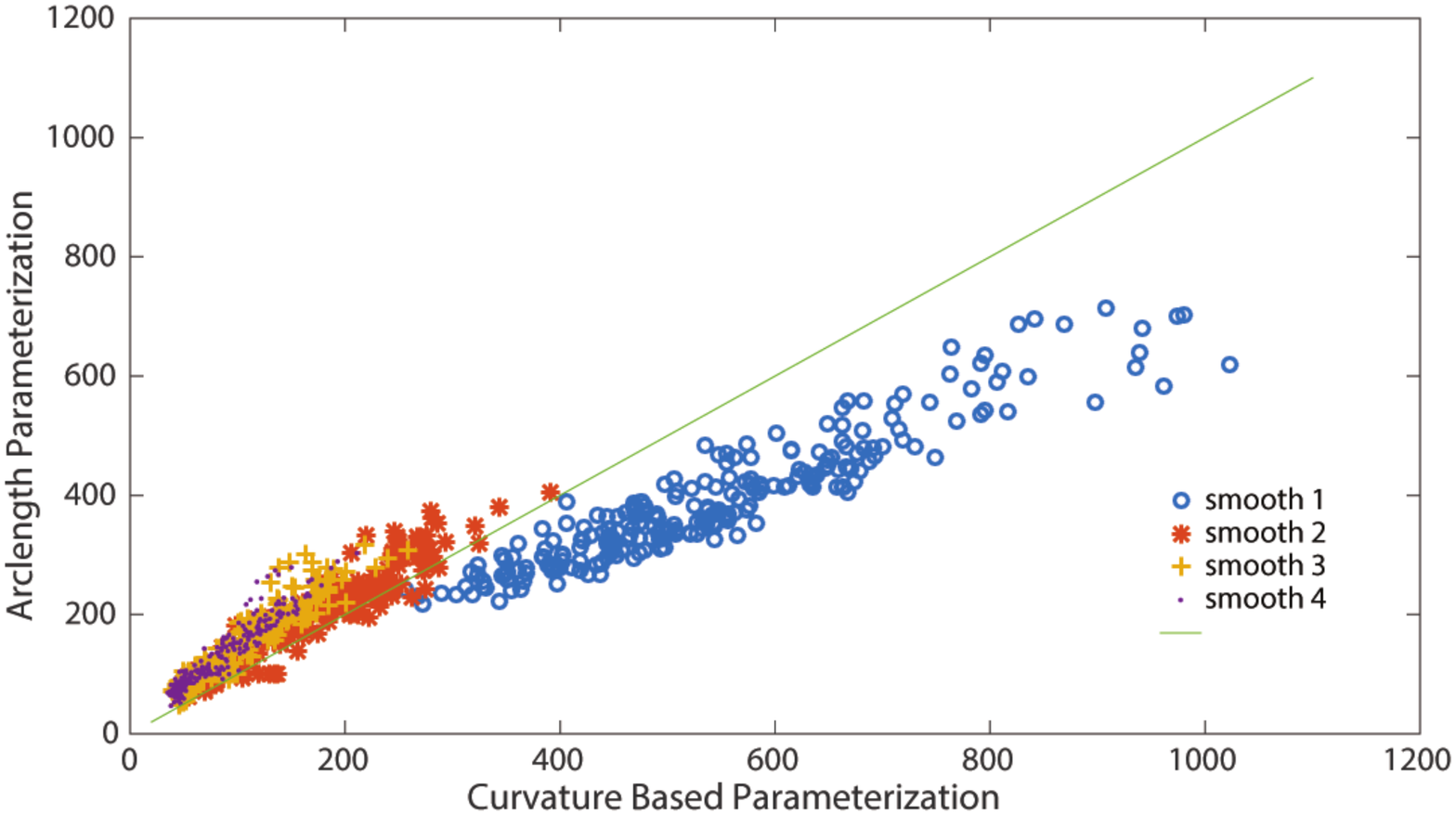}
    \includegraphics[scale=0.25]{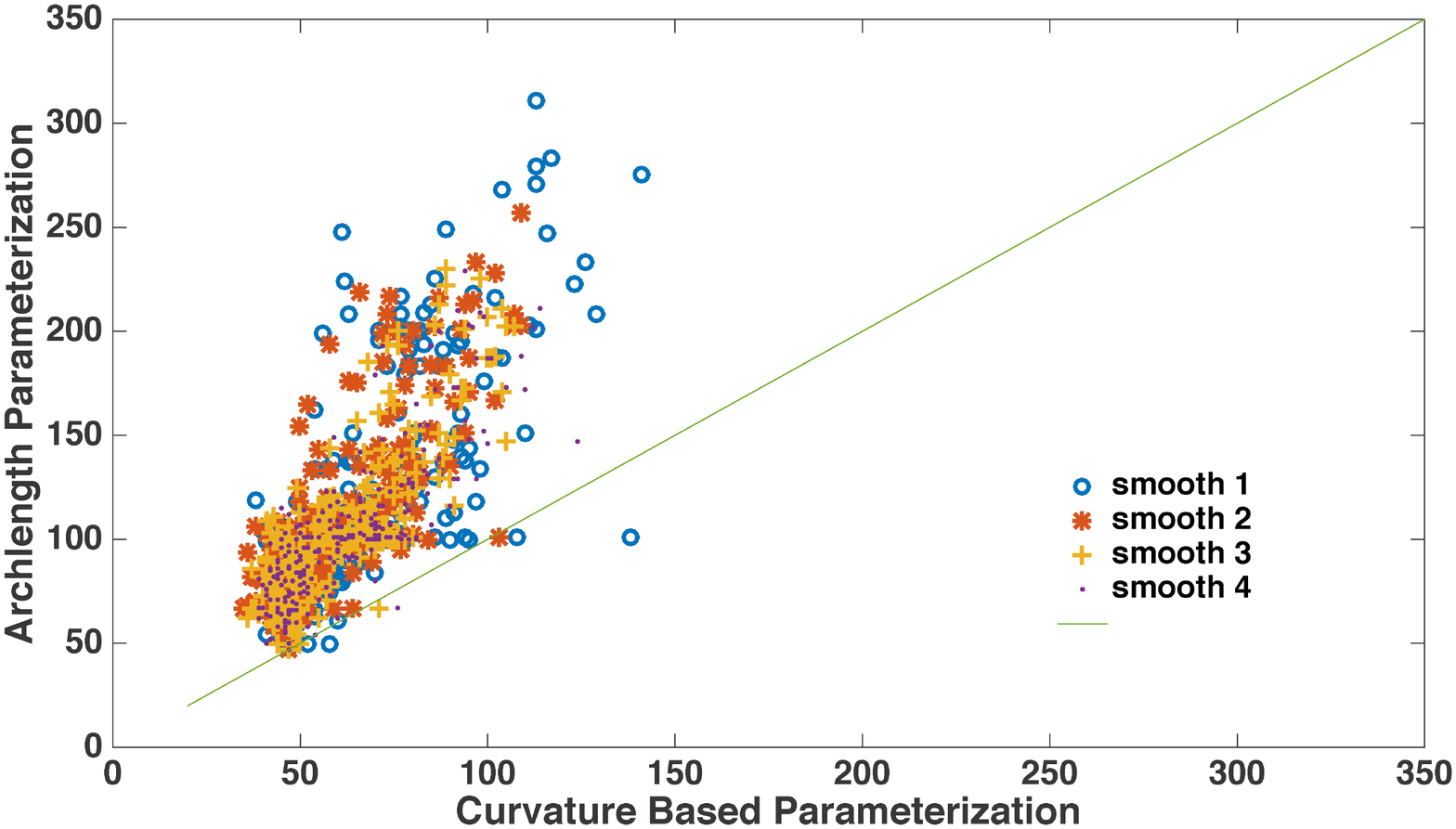}

	\caption{A comparison of $k_{LA}$ and $k_{LC}$ and $k_{DA}$ and $k_{DC}$, respectively, under various amounts of smoothing. \label{f:scattter-l}}
}
\end{figure}

To illustrate the impact of smoothing, we compared both $k_{LA}$ and $k_{LC}$ and $k_{DA}$ and $k_{DC}$ for the data set described in Section 7.1.  Intuitively, since the curvature parametrization explicitly relies on geometric features of the contours, we would expect that, for a given contour, $k_{LC}$ would typically be lower than $k_{LA}$ (with the same also holding for $k$ obtained through the distance criterion).  However, for data smoothed just once, we saw that this was typically not the case, as shown in Figure \ref{f:scattter-l}.  As an example of why this occurs, consider the example shown in Figure \ref{f:dog-smooth}. The small fluctuations along otherwise smooth regions cause the sampling points chosen according to curvature to be placed too close to each other under inadequate amounts of smoothing.

However, for the bounds obtained using the length criterion, the relationship between the two parameterizations appear to be resolved after subsequent uses of the moving average smoother.  Indeed, by its fourth use, it appears that the problem is fixed under both approximation criteria.  As such, for all subsequent data analysis, we utilize data that has been smoothed four times.  Please note that the relationship between the bounds is far less clear for the distance criterion due to the difference in shape being a highly non-linear feature, unlike the difference in length.

% just the models not the parameter estimates. need more data to that end.
\section{Predicting $k$}
\label{sec:pred}

Approximation of $k$ using the criteria discussed in Section \ref{sec:dc} is time consuming, especially if the distance criterion is used since those calculations are more complicated. The required time may range from several minutes to several hours depending on the variability in the curve and the number of points in the discretized observation. Therefore, we have developed regression models to predict $k$ using different characteristics of the original contour at a given error level, $E$. 

\subsection{Data and Predictors}

\begin{figure}[ht!]
	\center{
		\includegraphics[scale=0.5]{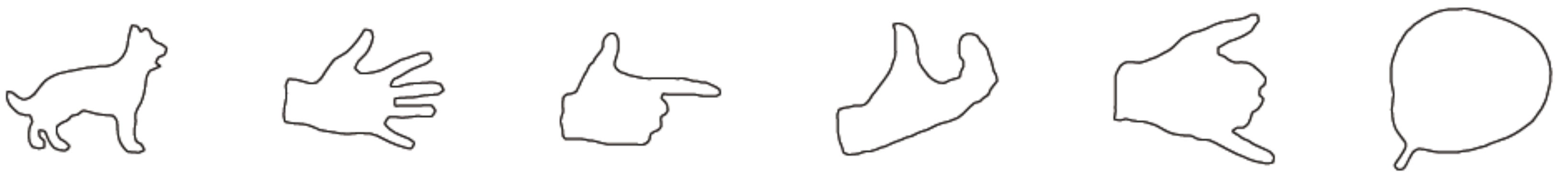}
		\includegraphics[scale=0.5]{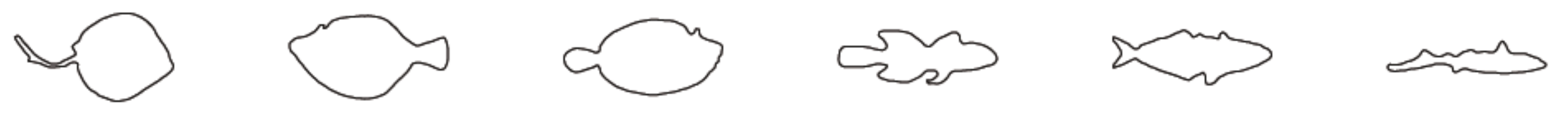}}
	\caption{ Representatives from the data set used for prediction.  \label{f:allfig}}
\end{figure}

We used the Kimia data set, which consists of 238 contours that are outlines of various objects. The first ten observations are of dogs. The next 120 observations are of six different species of fish, with 20 of each. There are also 20 contours of four different hand gestures (five each) and 88 different observations of pears.  Representatives from each class are shown in Figure \ref{f:allfig}.   

\subsection{Prediction of $k$ using Regression at error threshold $E$}

We fit linear regression models to predict the lower bound for the number of sampling points under each sampling scheme for both error criteria.  As such, we worked with four models, each having one of the following response variables: $k_{LA}, k_{DA}, k_{LC}  \mbox{ or } k_{DC}$.  Since the goal of these models is to simplify the process of finding lower bounds for $k$, we considered only predictors which are computationally simple to calculate from the original contours.

Our first predictor is $\left|\kappa\right|$, the total absolute curvature of the original contour.  The absolute curvature at a given point on the contour can be calculated as 
\begin{equation*}
\left|\kappa_i\right|=\frac{\left|x'y''-y'x''\right|}{\left(x'^2+y'^2\right)^{3/2}},
\end{equation*} 
for $i=1,\dots,K$, where $x$ and $y$ denote the real and imaginary coordinates, respectively, of the point.  As such, $\left|\kappa\right|$ is the sum of all of these local curvatures. The second predictor is the length of the original contour,
\begin{eqnarray*}
L_K=\sum_{j=2}^{K+1} \|z(t_j) -z(t_{j-1})\|
\end{eqnarray*} 
where $z(t_{K+1})=z(0)$ respectively. Our final predictor is $n_k$, the number of times the curvature's sign changes along the contour.  This is used as a surrogate for the number of ``features" the contour has and provides information not included in $\left|\kappa\right|$ since it does not consider signs.  It should be noted that the total curvature, itself, would provide no useful information since that quantity is always zero for any closed curve. The resulting regression equation for predicting $k$ at a given, $E$ will take the form of
$$k=\beta_0+\beta_1 \left|\kappa\right|+\beta_2 L_k+\beta_3 n_k+\epsilon$$

We first calculated the values of the predictors for every observation and chose to use $E=0.005$ as the approximation threshold.  This is the same threshold that was used to generate Figures  \ref{f:scattter-l} and \ref{f:scattter-d} and was chosen since, even though it requires extremely close approximations, the resulting dimension reduction still results in $k<<K$.  We then approximated $k_{LA}$, $k_{DA}$, $k_{LC}$ and $k_{DC}$ at this level for all $238$ observations in the data set using the method discussed in Section \ref{sec:dc} and performed model selection.  The significant predictors and their parameter estimates are listed in Tables \ref{t:para1} and \ref{t:para2} along with with the corresponding p-values. We examined the residuals for these models using graphical tests and verified that no model assumptions were violated.

\begin{table}[ht!]
	\centering
	\caption{Parameter Estimates and statistics  for Predicting $k$ under arc length parameterization at $E=0.005$\label{t:para1}}
	\begin{tabular}{l|rrrrrr}
		\hline
		Decision Criteria & \multicolumn{3}{c|}{Length $\left( k_{LA}\right)$} & \multicolumn{3}{c}{Distance $\left( k_{DA}\right) $} \\
		& Estimate & & \multicolumn{1}{c|}{p-value}    & Estimate &  & p-value \\
		\hline
		Intercept & $-20.8966$ & &	\multicolumn{1}{c|}{$9.76\times 10^{-5}$}	& $-18.1966$	& & $0.0129$ \\
		$|\kappa|$ & $5.4190$ &	& \multicolumn{1}{c|}{$2\times 10^{-16}$}	& $2.5298$ & &	$2\times 10^{-16}$ \\
		$L_K$ & $-0.0264$ & &	\multicolumn{1}{c|}{$9.09\times 10^{-5}$}	& $0.0638$ &	& $8.12\times 10^{-5}$ \\
		$n_\kappa$ & & & \multicolumn{1}{c|}{} &  $-0.1229$ & &	$0.0411$ \\	 	 
		%15.05	235	18.76	234	7.865	234	25.98	236
		$RMSE$ & $14.9548$	 & &	\multicolumn{1}{c|}{} &	$18.6017$ & & \\
		%0.9293	 	0.7296	 	0.9575	 	0.5228	 
		$F$ & $1558$ & &	\multicolumn{1}{c|}{$2.2\times10^{-16}$} &	$214.1$	& & $2.2\times10^{-16}$ \\
		\hline
	\end{tabular}%
	\label{t:kl}%
\end{table}

\begin{table}[ht!]
	\centering
	\caption{Parameter Estimates and statistics  for Predicting $k$ under curvature parameterization at $E=0.005$ \label{t:para2}}
	\begin{tabular}{l|rrrrrr}
		\hline
		Decision Criteria & \multicolumn{3}{c|}{Length $\left( k_{LC}\right) $} & \multicolumn{3}{c}{Distance $\left( k_{DC}\right) $} \\
		& Estimate & & \multicolumn{1}{c|}{p-value}    & Estimate &  & p-value \\
		\hline
		Intercept &  $8.4779$	& & \multicolumn{1}{c|}{$0.0343$} &	$9.3872$ &	& $6.96 \times 10^{-5}$ \\
		$|\kappa|$ &  $3.7658$	& & \multicolumn{1}{c|}{$2.2\times 10^{-16}$} &	$1.5302$	& & $1.02\times 10^{-90}$ \\
		$L_K$ &	$-0.0258$ &	& \multicolumn{1}{c|}{$1.43\times 10^{-5}$}  & $0.0162$ & & $1.5\times 10^{-3}$\\
		$n_\kappa$ & $-0.1324$ & &	\multicolumn{1}{c|}{$2.99\times 10^{-7}$} & $-0.0709$ & &  $2.59\times 10^{-4}$\\	 	 	%15.05	235	18.76	234	7.865	234	25.98	236
		$RMSE$ & $7.7986$ 	 & &	\multicolumn{1}{c|}{} & $5.99$ 	& &	\\
		$F$ &	$1780$ & & \multicolumn{1}{c|}{$2.2\times 10^{-16}$}	& $616$	& & $9.99\times10^{-111}$ \\
		\hline
	\end{tabular}
	\label{t:kl}
\end{table}

Note that $n_k$ does not have a significant effect on $k_{LA}$ and the only significant predictor in the model for predicting $k_{DC}$ is $\left| \kappa\right|$. For $k_{DA}$ and $k_{LC}$, though, the full models were found to be significant. The Root Mean Squared Error (RMSE) is also provided in each table. 
%The model for $k_{LC}$ has the smallest RMSE, providing some initial evidence that it provides the best fit among the four, and the model for $k_{DC}$ has the highest RMSE, suggesting that that model is less adequate.  
{The models for $k_{LC}$ and $k_{DC}$ have smaller RMSE, providing some initial evidence that curvature parameterization provide  best fits among the two parameterizations, and the model for $k_{DA}$ has the highest RMSE.} However, since the highest priority for these models is their use in predicting the lower bounds, we used cross validation to examine the predictive performance of the models.

\subsection{Model Validation}

We use two different methods for validation. The first method we used is $80: 20$ cross validation to examine the predictive performance for new data similar to the training data set.  To examine the predictive performance for new classes of data dissimilar to the training set, we used leave-one-category out cross validation.

\subsubsection{$80:20$ cross validation}
\label{ssec:cv1}

\begin{figure}[ht!]
	\center{
		\subfigure[\footnotesize{$\mbox{ mean } \approx15 \mbox{ minimum } \approx 8 \mbox{ range } \approx 14$}]{\includegraphics[scale=0.21]{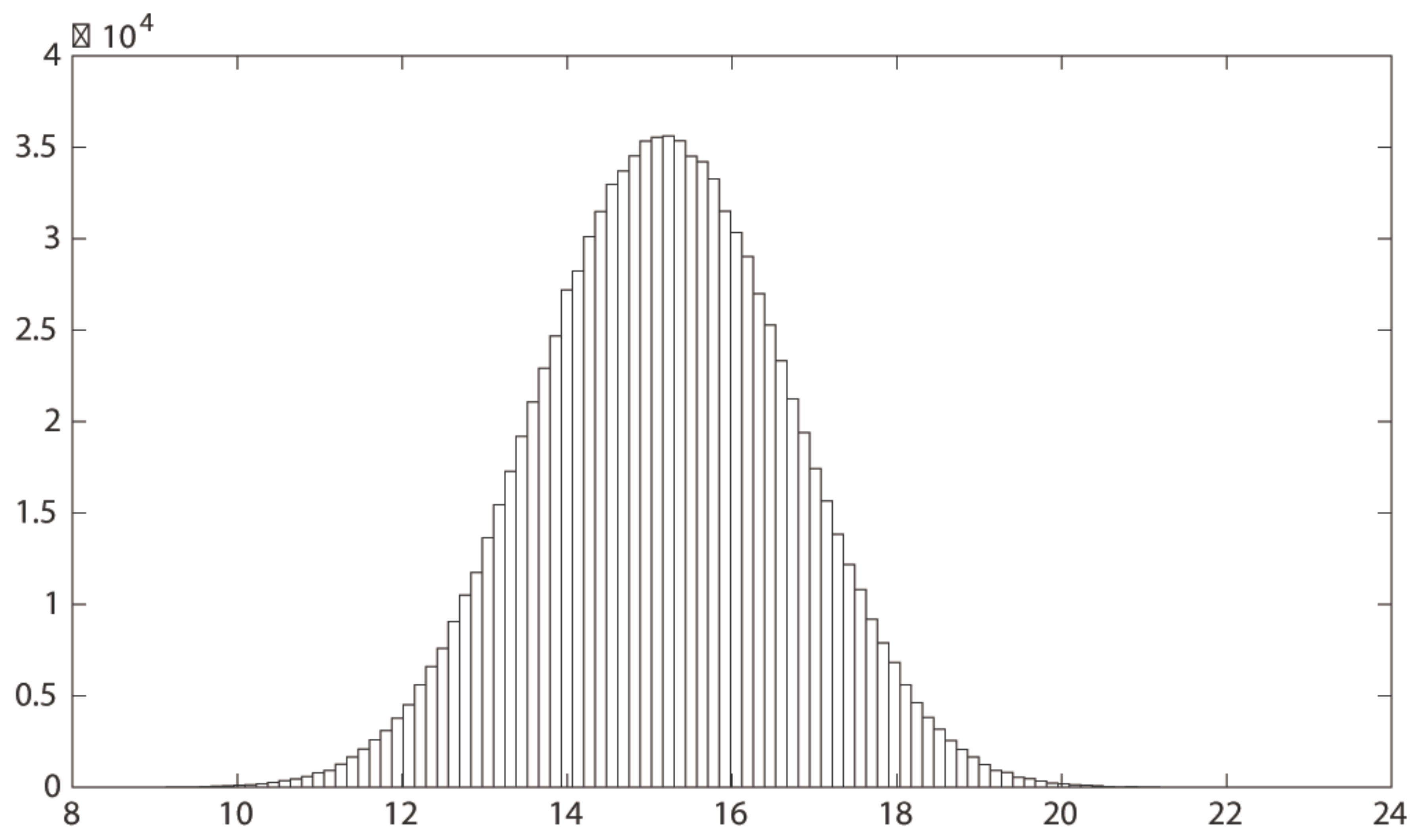}} 
		\subfigure[\footnotesize{$\mbox{ mean } \approx 19 \mbox{ minimum } \approx 11 \mbox{ range } \approx 16$}]{\includegraphics[scale=0.21]{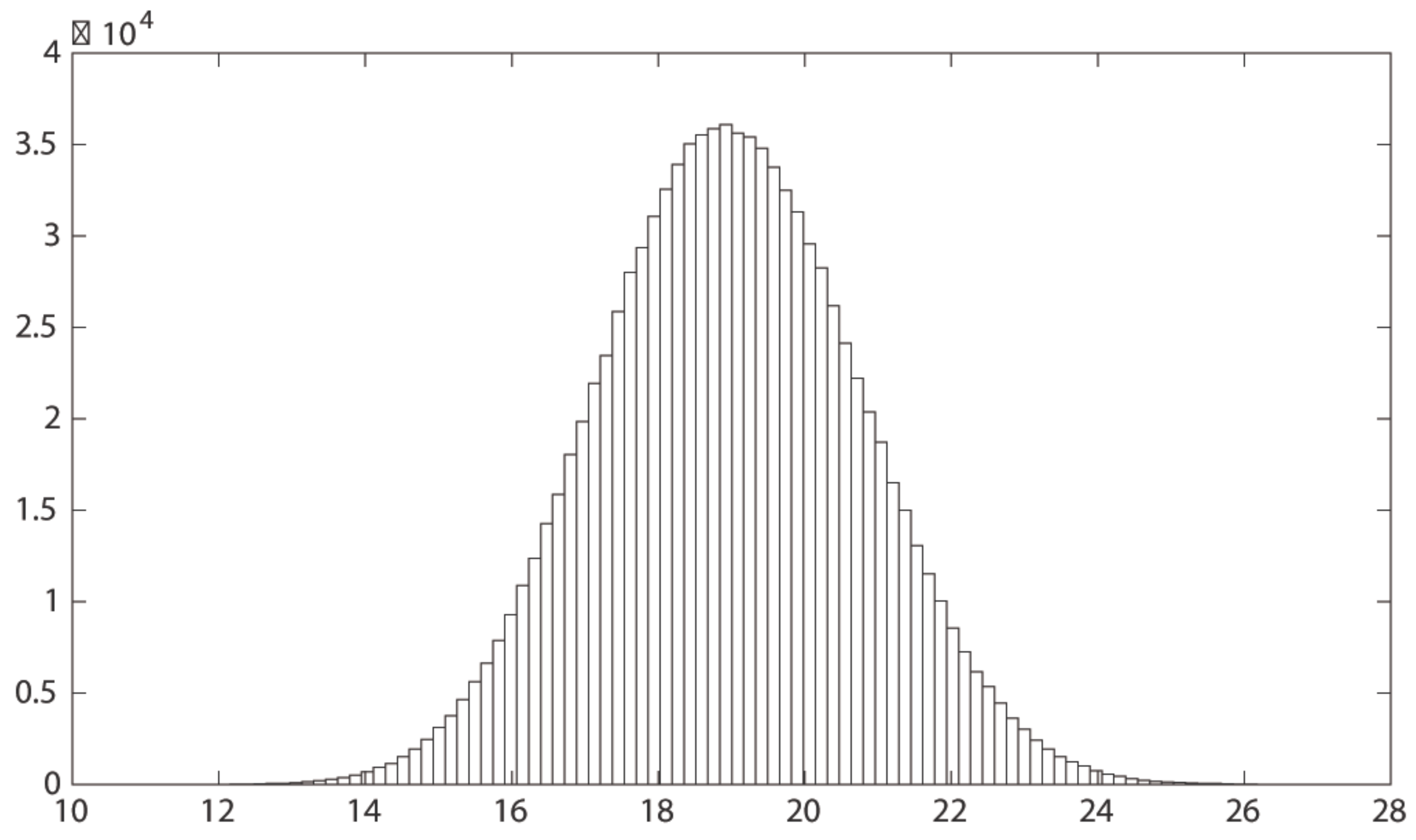}}\\
		\subfigure[\footnotesize{$\mbox{ mean } \approx 8 \mbox{ minimum } \approx 4 \mbox{ range } \approx 10$}]{ \includegraphics[scale=0.21]{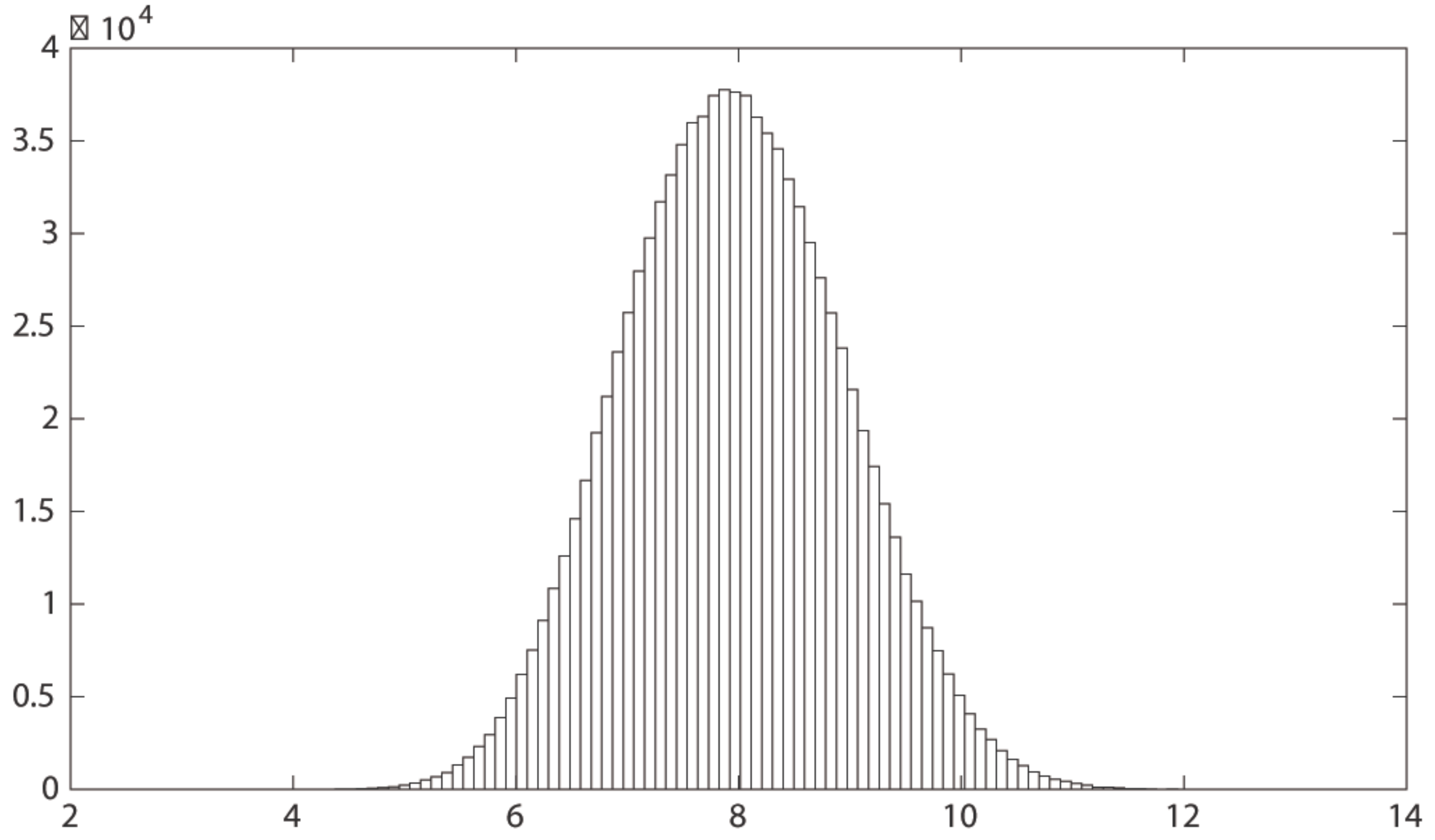}}
		\subfigure[\footnotesize{$\mbox{ mean } \approx 6 \mbox{ minimum } \approx 3 \mbox{ range } \approx 6$}]{ \includegraphics[scale=0.21]{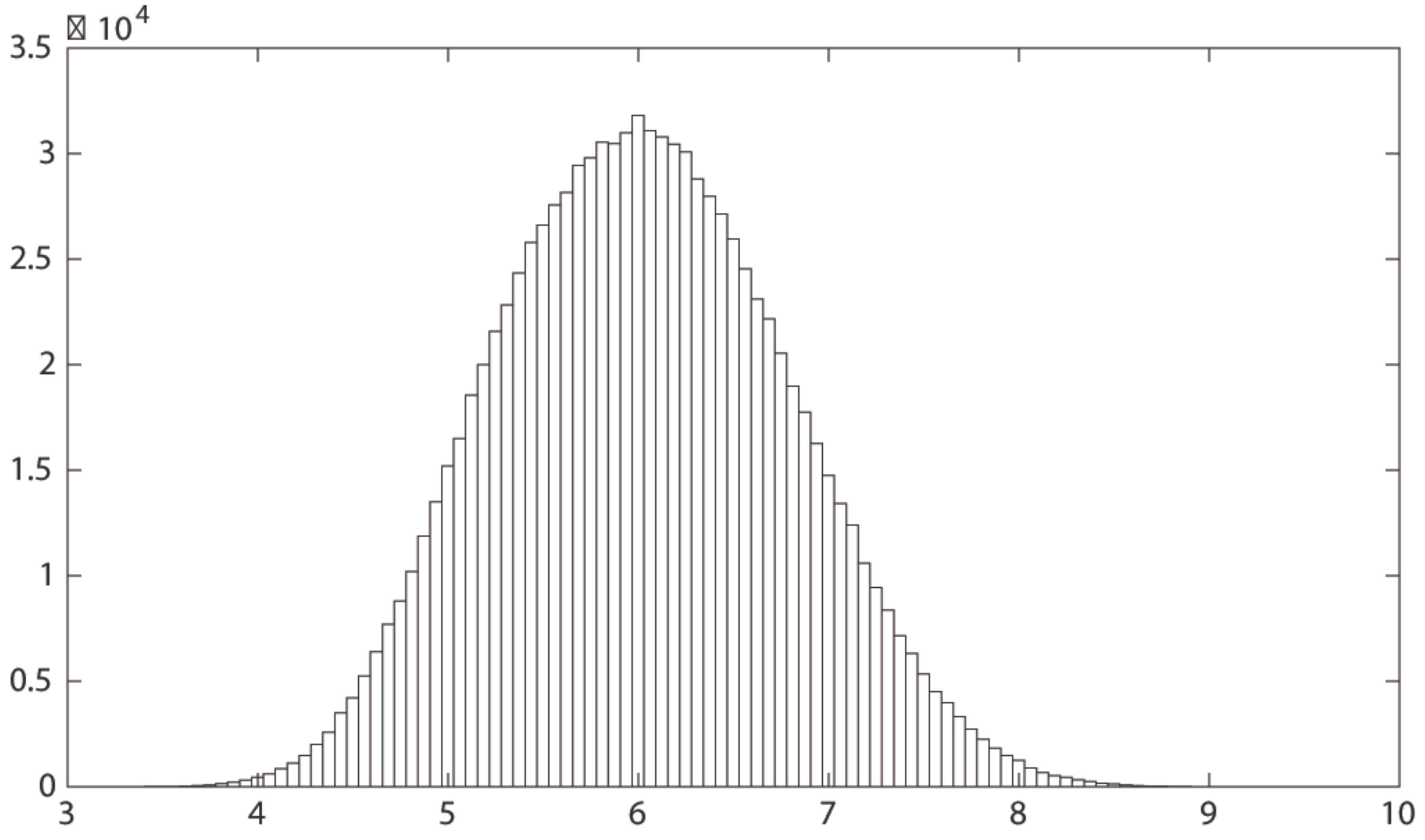}}}\\
	\caption{Probability distributions of RMSE for one million test sets under $80\% $ to $20\%$ cross validation. (a) $k_{LA}$, (b) $k_{DA}$, (c) $k_{LC}$, and (d) $k_{DC}$ \label{f:hist8020}}	
\end{figure}

First, we randomly divided the data set in to training and test sets so that  $80\%$ of the observations were in the training set and the remaining $20\%$ were in the test set. Using the data in the training set, we estimated the model parameters for all four final models. Then we used those parameter estimates to predict the responses in the test set and find the RMSE of the predicted values.  We repeated this process $1,000,000$ times. Histograms of the distributions of RMSE for the four models are shown in Figure \ref{f:hist8020}. 

The most noticeable feature of the histograms is that the distributions of RMSE appear to be approximately normally distributed. Secondly, we see that the mean RMSE for each distribution is comparable to the observed RMSEs for the full data set, despite being based on fewer observations.  This suggests that the models are able to predict lower bounds for $k$ for contours similar to those used to form the models quite well.  Based on the means and ranges of these distributions, the models can be listed in increasing order of predictive performance as, $k_{DA}, k_{LA}$, $k_{LC}$ and  $k_{DC}$. 

\begin{figure}[ht!]
	\center{
		\subfigure[\footnotesize{$\mbox{ mean } \approx14 \mbox{ minimum } \approx 1 \mbox{ range } \approx 39$}]{\includegraphics[scale=0.21]{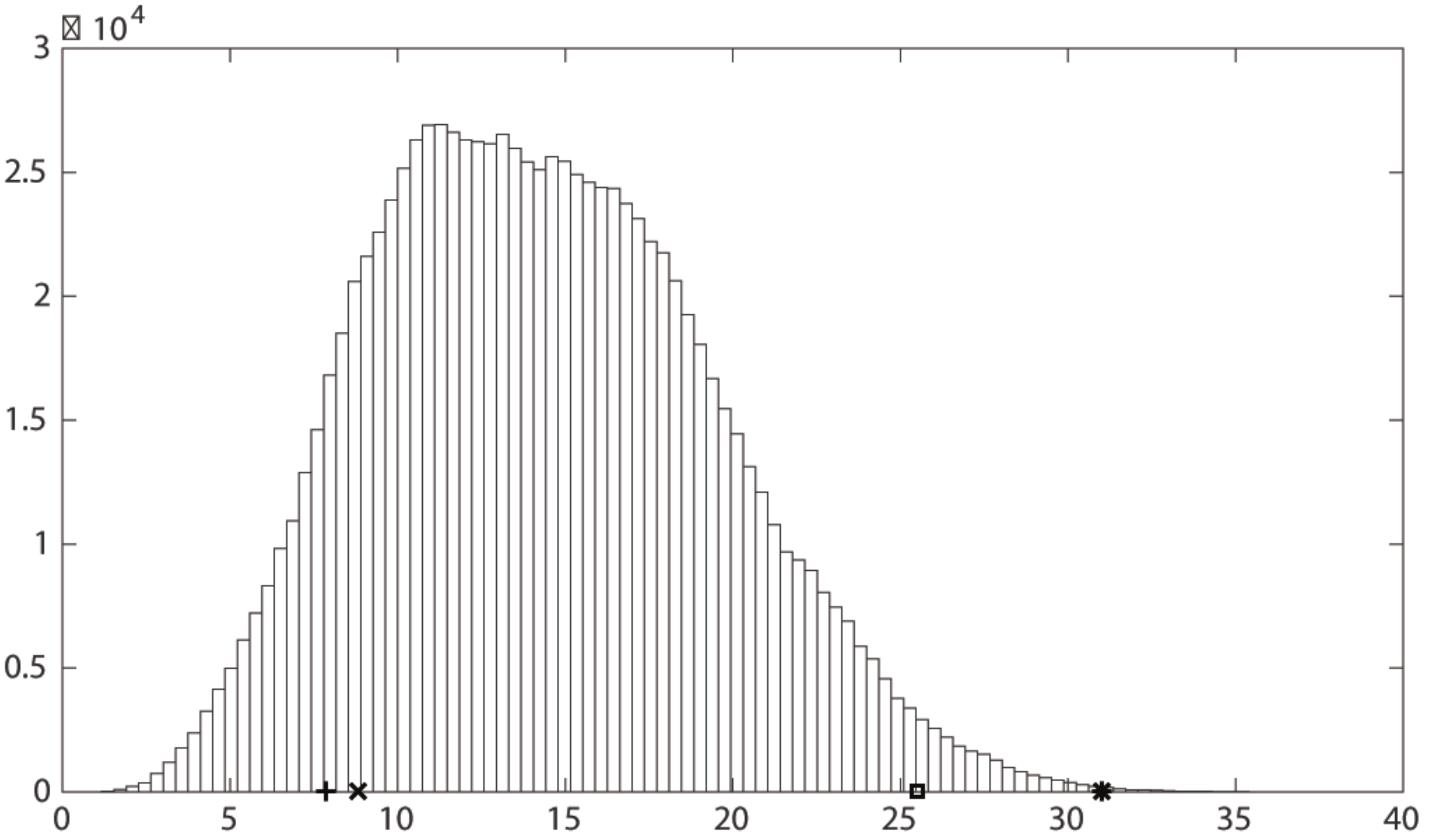}} 
		\subfigure[\footnotesize{$\mbox{ mean } \approx 18 \mbox{ minimum } \approx 0 \mbox{ range } \approx 47$}]{\includegraphics[scale=0.21]{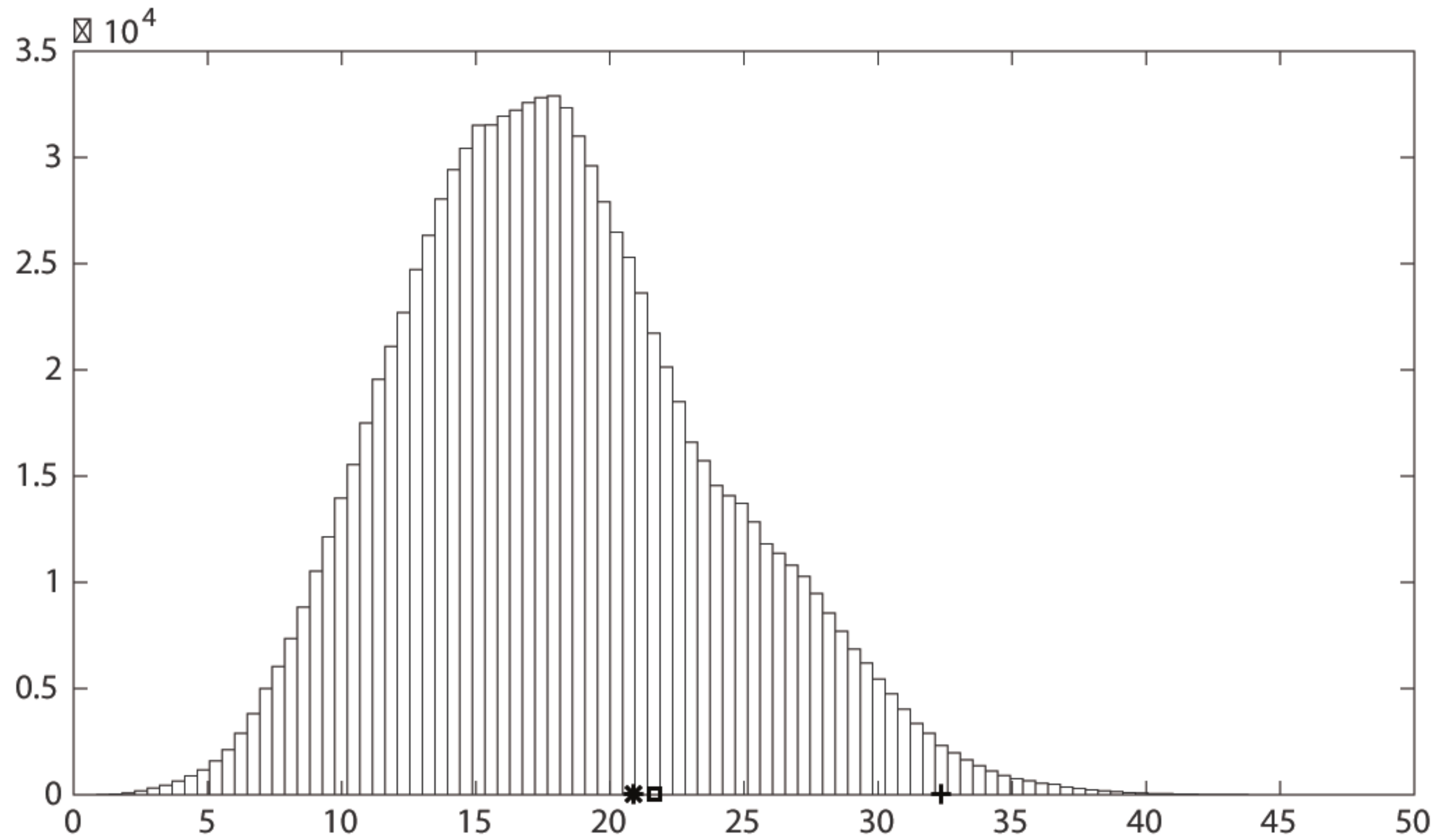}}\\
		\subfigure[\footnotesize{$\mbox{ mean } \approx 7 \mbox{ minimum } \approx 0 \mbox{ range } \approx 23$}]{ \includegraphics[scale=0.21]{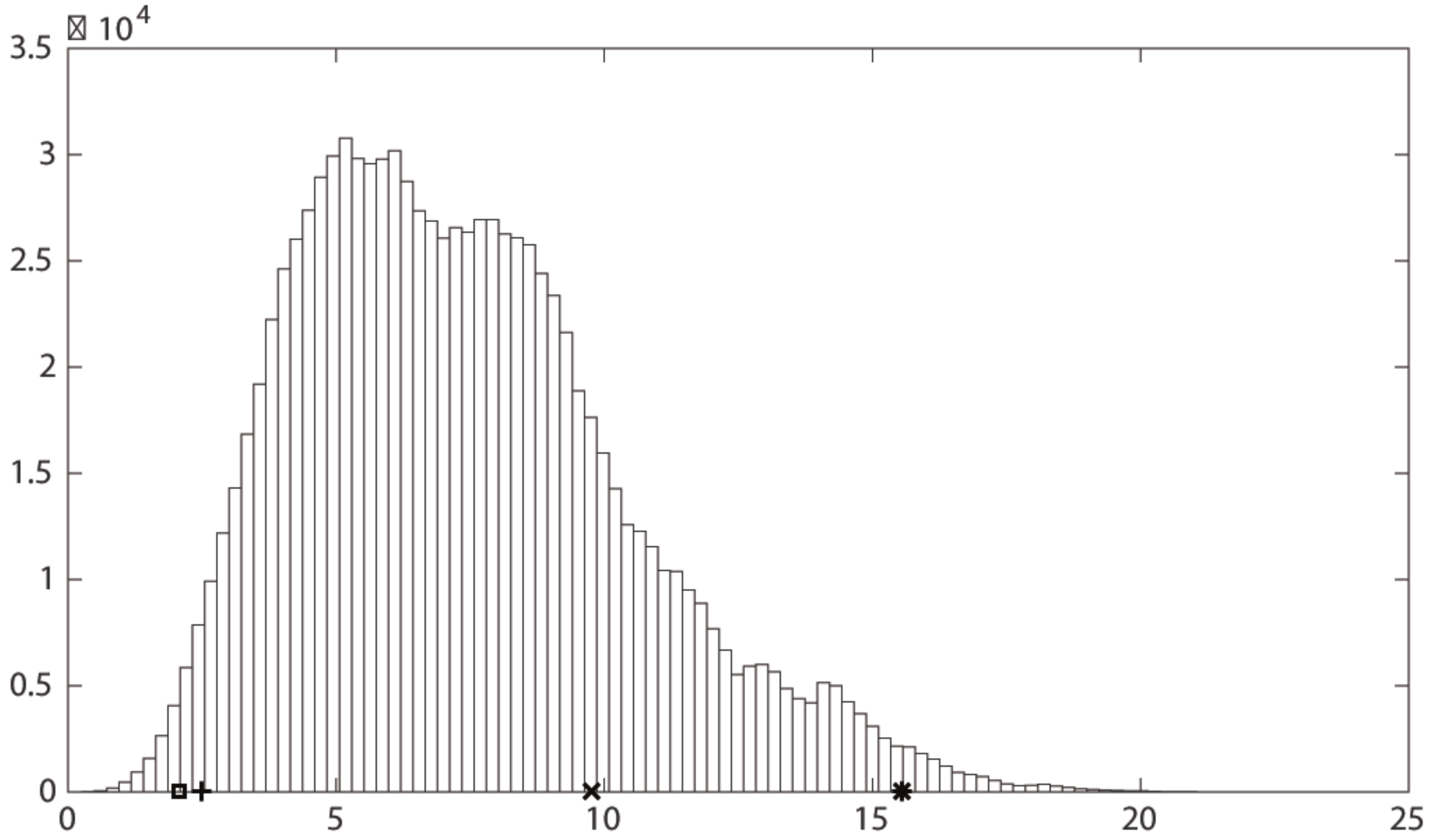}}
		\subfigure[\footnotesize{$\mbox{ mean } \approx 8 \mbox{ minimum } \approx 0 \mbox{ range } \approx 24$}]{ \includegraphics[scale=0.21]{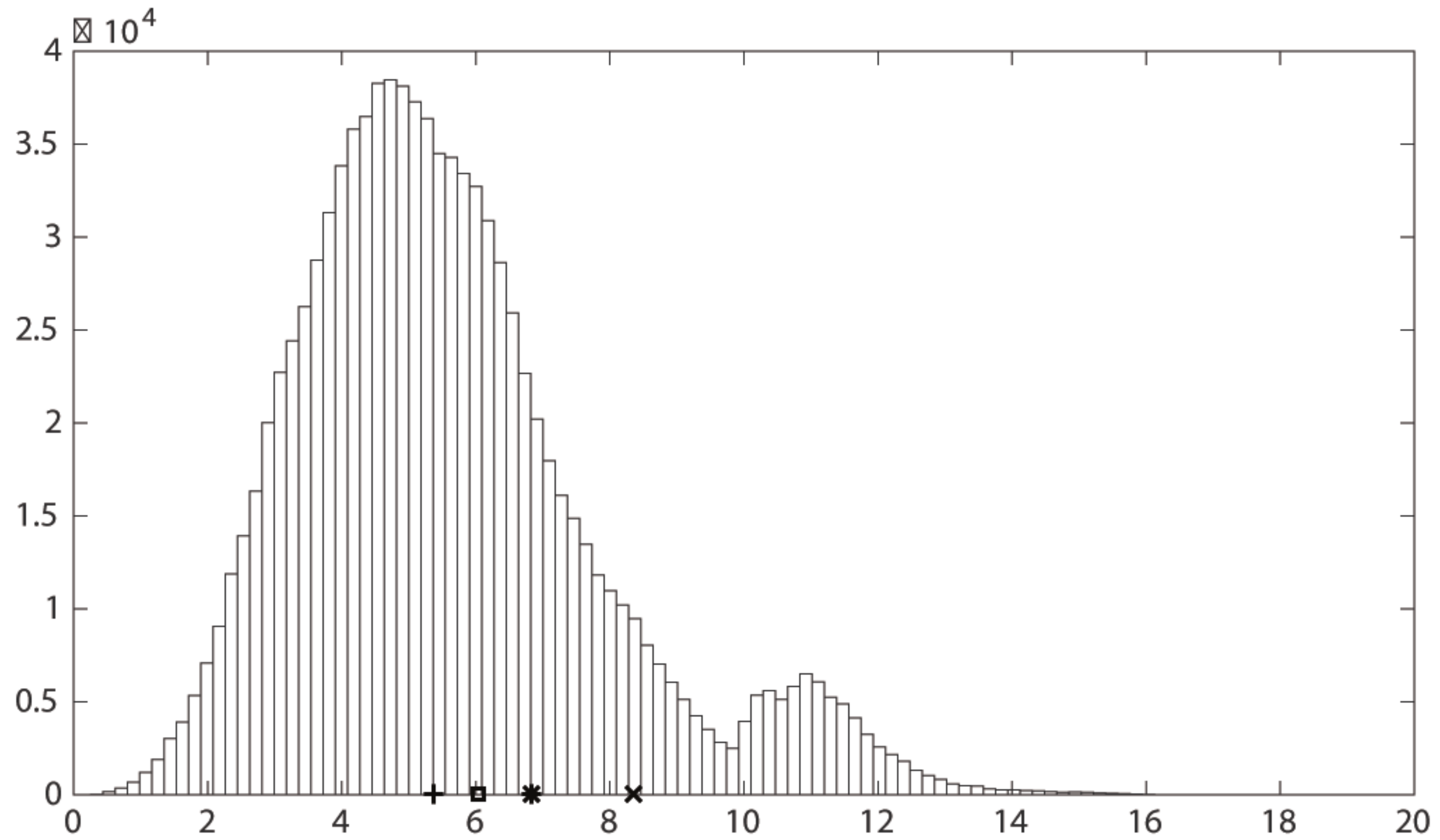}}}\\
	\caption{Histograms for test sets of size $5$ compared to four categories of hand gestures. (a) $k_{LA}$ (b) $k_{DA}$ (c) $k_{LC},$ and (d) $k_{DC}$. \label{f:rmseh}}
\end{figure}

\subsubsection{Leave one-category out cross validation}
\label{ssec:cv2}

To examine data dissimilar to that which is used to obtain the models, we then removed an entire category of observations at a time to use as our training set.  We did this for each of the 12 categories in the data set.  The RMSE for each class is shown in Table \ref{t:valid1}. To assess predictive performance, though, we needed to compare these observed values to distributions of RMSEs obtained by, again, randomly removing observations for the test set.  However, since the difference classes of data contain different numbers of observations, most which are considerably different from the $80:20$ ratio, we needed to repeat this process four times using test sets of size 5, 10, 22, and 88, for the hand gestures, dogs, fish, and pears, respectively.  For each class, instead of generating all possible samples of that given sample size to construct the exact distribution, we approximated the distributions by again taking $1,000,000$ test sets of each size.

\begin{table}[ht!]	
	\centering
	\caption{Predictive performance for each category of data when used as the test set \label{t:valid1}}
	\small{
		\begin{tabular}{l|r|r|r|r|r|r|r|r}
			&   \multicolumn{4}{c|}{$RMSE$ by category} & \multicolumn{4}{c}{Percentage of positive residuals} \\
%			\cmidrule{2-5} \cmidrule{6-9}
			Type & $k_{LA}$ & $k_{DA}$ & $k_{LC}$ & $k_{DC}$ & $k_{LA}$ & $k_{DA}$ & $k_{LC}$ & $k_{DC}$\\
			\hline
			Dog	& 	24.67	&	28.07	&	15.36	&	6.79	& 20  &  70 &  80  &  40\\
			Fish	$\#1$ & 	20.76	&	23.32	&	11.58	&	11.05 &  35  &   50 &  30  &  75\\
			Fish	$\#2$	& 	12.22	&	13.42	&	7.45	&	6.31 &  40  &  40 &  25  &  35 \\
			Fish	$\#3$	& 	13.37	&	15.04 &	4.68		&	5.64 &  45  &   25 &  45  &  20	\\
			Fish	$\#4$	&	14.08	&	13.98 &	7.95		&	5.84 &  55  &  45 &  55 &   40\\
			Fish	$\#5$	& 	20.93	&	21.08 &	12.04	&	9.44 &  80  &  50 &  75   &  65\\
			Fish	$\#6$	& 	15.51	&	23.48 &	9.32	&	6.05 & 35 &  25 &   60 &  20\\
			Hand	$\#1$	& 	31.00	&	20.90 &	15.56	&	6.83 & 100   &  100 & 0  &  80\\
			Hand	$\#2$	& 	8.81	&	20.88&	9.77	&	8.36	& 40  &  20  &  20 & 40\\
			Hand	$\#3$	& 	25.49	&	21.69 &	2.07	&	6.03 & 20  & 0  &  40  & 0\\ 
			Hand	$\#4$	& 	7.86	&	32.33 &	2.50	&	5.37 & 80  &  20  &  80  &  20\\
			Pear		& 	12.55	&	22.44 &	9.16	&	4.78 & 75 &   88  & 86 &  81\\
			\hline
		\end{tabular}}
	\end{table}	

To assess the predictive performance of each model, we first calculated the proportion of RMSEs above the observed value for every class of data based upon the idea that higher proportions suggest good predictive performance.  These upper tail probabilities are shown in Table \ref{t:valid2}.  While many of these proportions are quite high, some are certainly quite low, as with Fish $\#1$.  While this seems to suggest poor predictive performance, it is also important to consider whether the predicted values are greater or less than the observed values of $k$.  Since the response variable is a lower bound, overestimation is not nearly as undesirable as under-prediction.  Therefore, to understand this, we calculated the percentage of positive residuals for each category.  These are also shown in Table  \ref{t:valid1}.

	\begin{table}[ht!]	
		\centering
		\caption{Upper tail probability for RMSE of each category for each model \label{t:valid2}}
		\small{
			\begin{tabular}{l|r|r|r|r}
				Type &  $k_{LA}$ & $k_{DA}$ & $k_{LC}$ & $k_{DC}$ \\
%				\midrule
				Dog		& 0.0036 & 0.0153 & 0.0008 & 0.2462	\\
				Fish	$\#1$	& 0.0114 & 0.0656 & 0.0130 	& 0.0000\\
				Fish	$\#2$	& 0.8651 & 0.9719 & 0.5728 & 0.3554	\\
				Fish	$\#3$	& 0.7397 & 0.9005  & 0.9851 & 0.5473 	\\
				Fish	$\#4$	& 0.6395 & 0.9545 & 0.4513 & 0.4857	\\
				Fish	$\#5$	& 0.0095 & 0.2188	& 0.0060 & 0.0037 	\\
				Fish	$\#6$	& 0.4188 & 0.0590 & 0.1825 & 0.4246 	\\
				Hand	$\#1$	& 0.0007 & 0.2903 & 0.0123 & 0.2324 	\\
				Hand	$\#2$	& 0.8602 & 0.2915 & 0.2015 & 0.1139  \\
				Hand	$\#3$	& 0.0189 & 0.2517 & 0.9905 & 0.3529	\\
				Hand	$\#4$	& 0.9080 & 0.0124 & 0.9782 & 0.4755	\\
				Pear		& 0.9959 & 0.0027 & 0.0433 & 0.9954	\\
				\hline
			\end{tabular}}
		\end{table}

As an illustrative example, we will provide an in depth analysis for the four categories of hand gestures.  Histograms for the distributions of RMSE when using test sets consisting of 5 observations are shown in Figure \ref{f:rmseh}.  Please note that the observed RMSE for each class of hand gesture is marked on each histogram.  The symbols, $\ast, \times, \square, \; \mbox{and}, \; +$ represent the RMSEs for hand gestures $\#1, \#2, \#3, \mbox{and}, \;\; \#4$ respectively. We see that the models for $k_{LA}$, $k_{LC}$, and $k_{DN}$ do not predict Hand $\#1$ well. In each case the upper tail probability is quite low, though the observations were frequently overestimated. It is not particularly surprising that this class tends to have higher RMSEs, though, since, as shown in Figure \ref{f:allfig}, this type of contour has considerably more features and is thus considerably different than the others. We also see that the predictive performances of $k_{LA}, k_{LC}$ and $k_{DN}$ for Hand $\#4$ are very good having very high upper tail probabilities and percentages of positive residuals. The model for $k_{LC}$ is very good in predicting Hand $\#3$, having a very hight right tail probability. Similarly, the model for $k_{DN}$ is the best among four models in predicting Hand $\#2$. 

Overall, we see that the Fish $\#1$ is the only other class generally not well predicted across the board.  While some other classes, such as Fish $\#5$, have small upper tail probabilities, the high RMSEs are typically the result of overestimation.  Just as with the hand gestures, we see that the $k_{DC}$ model generally produces the smallest RMSE values for all categories, suggesting that it has the best predictive performance of all of them.  Indeed, the only class it does a particular poor job with is Fish $\#1$.  Across the board, though, we see that these models are able to predict a lower bound for $k$ fairly well when faced with new categories of observations except for classes that are considerably different than those used to build the model.

\section{Discussion}

In this paper, we presented a framework for determining lower bounds for the number $k$ of sampling points needed for approximating planar configurations for shape analysis.  We considered two criteria for doing this:  the relative error in the length of the $k$-gon and due to approximation and the relative distance between he $k$-gon and the original contour.  The latter criterion tends to require fewer sampling points than the former, but tends to result in the approximation not capturing fine details of the contours.  This leads to the question of whether it is more important for an analysis to approximate the contour, which the length criterion evaluates, or the actual shape of the contour, which the distance criterion evaluates.  Ultimately, the answer to this question is likely to be application dependent.  It may also be possible to find additional criteria that may balance the approximation of the contour and its shape.

For each criterion, we considered two different sampling point selection methods, each corresponding to a different parametrization for the contour.  The equally spaced sampling points are based upon an arc length parametrization of the contour and polygons obtained from them tend to approximate the overall form of the contour well, but are more prone to not capturing small details.  On the other hand, the sampling points that retain equal amounts of absolute curvature between them are based on a curvature parametrization of the contour.  Polygons obtained using these tend to capture the small details of the contours better.  However, complications may arise using these for subsequent analyses because it is important for shape analysis that sampling points to correspond across observations.  The quality of the correspondences may not be as good with this parametrization. Other sampling point selection methods could also be used, such as the random selection method of Ellingson et al. (2013).  However, due to the random nature of that approach, it will be more difficult to determine a lower bound for $k$ since the lengths and distances will be random.  In all of these cases, though, it is important to adequately smooth the data first due to fluctuations in the contours resulting from digitization.

While this framework is effective at reducing the dimensionality of the data, it is computationally inefficient.  To help remedy this situation, we developed regression models for predicting the lower bounds for $k$ by using predictors that are computationally cheap to calculate.  Cross validation studies suggest that these models are able to predict these bounds relatively well for new data, regardless of whether the new observations are particularly similar to the data used to fit the model.  Furthermore, it appears that the model for predicting the lower bound for $k$ using the length criterion when the sampling points are chosen according to curvature is especially effective.  We should note, though, that here we considered only the approximation threshold of $0.005$.  More work remains to be done to see how the models vary across values of $E$ and how this value impacts predictive performance.  Additionally, it remains to be seen how this error threshold will impact any subsequent shape analyses, including the calculation of mean shapes and discrimination between categories of contours.

%\begin{center}BIBLIOGRAPHY\end{center}

 \end{document}